\documentclass[twocolumn]{emulateapj}

\usepackage{graphicx}
\usepackage{hyperref}
\usepackage{amssymb}
\usepackage{amsmath}

\def\aap {{A\&A}}

\def\apjl {{ApJL}}

\def\apjs {{ApJS}}

\newcommand{\FLASH}{{\sc flash} }
\newcommand{\HYPRE}{{\sc hypre} }
\newcommand{\PLUTO}{{\sc pluto} }
\newcommand{\MAKEMAKE}{{\sc makemake} }
\newcommand{\HII}{H{\sc ii} }
\newcommand{\Msun}{M_\odot}

\newcommand{\Rsun}{R_\odot}
\renewcommand{\vec}[1]{\boldsymbol{#1}}
\newcommand{\grad}{\vec{\nabla}}
\renewcommand{\div}{\grad \vec{\cdot} }

\shorttitle{Hybrid radiation transport on an adaptive grid}
\shortauthors{Klassen et al.}

\begin{document}
\bibliographystyle{apj}

\title{A general hybrid radiation transport scheme for star formation simulations on an adaptive grid}

\author{Mikhail Klassen}
\affil{Department of Physics and Astronomy, McMaster University \\ 1280 Main St.~W, Hamilton, ON L8S 4M1, Canada}
\email{klassm@mcmaster.ca}

\author{Rolf Kuiper} %\altaffilmark{1}}
\affil{Max Planck Institute for Astronomy \\ K\"{o}nigstuhl 17, D-69117 Heidelberg, Germany}

\author{Ralph E.~Pudritz\altaffilmark{1}}
\affil{Department of Physics and Astronomy, McMaster University \\ 1280 Main St.~W, Hamilton, ON L8S 4M1, Canada}

\author{Thomas Peters}
\affil{Institut f\"{u}r Computergest\"{u}tzte Wissenschaften, Universit\"{a}t Z\"{u}rich \\ Winterthurerstrasse 190, CH-8057 Z\"{u}rich, Switzerland}

\author{Robi Banerjee}
\affil{Hamburger Sternwarte, Universit\"{a}t Hamburg \\ Gojenbergsweg 112, D-21029 Hamburg, Germany}

\and

\author{Lars Buntemeyer}
\affil{Hamburger Sternwarte, Universit\"{a}t Hamburg \\ Gojenbergsweg 112, D-21029 Hamburg, Germany}

%\altaffiltext{1}{Max Planck Institute for Astronomy Heidelberg, K\"{o}nigstuhl 17, D-69117 Heidelberg, Germany}
\altaffiltext{1}{Origins Institute, McMaster University, 1280 Main St.~W, Hamilton, ON L8S 4M1, Canada}

\begin{abstract}
Radiation feedback plays a crucial role in the process of star formation. In order to simulate the thermodynamic evolution of disks, filaments, and the molecular gas surrounding clusters of young stars, we require an efficient and accurate method for solving the radiation transfer problem. We describe the implementation of a hybrid radiation transport scheme in the adaptive grid-based \FLASH general magnetohydrodynamics code. The hybrid scheme splits the radiative transport problem into a raytracing step and a diffusion step. The raytracer captures the first absorption event, as stars irradiate their environments, while the evolution of the diffuse component of the radiation field is handled by a flux-limited diffusion (FLD) solver. We demonstrate the accuracy of our method through a variety of benchmark tests including the irradiation of a static disk, subcritical and supercritical radiative shocks, and thermal energy equilibration. We also demonstrate the capability of our method for casting shadows and calculating gas and dust temperatures in the presence of multiple stellar sources. Our method enables radiation-hydrodynamic studies of young stellar objects, protostellar disks, and clustered star formation in magnetized, filamentary environments.
\end{abstract}


% Introduction
\section{Introduction} \label{sec:intro}

The last several years have seen increasingly sophisticated radiation feedback models implemented in a wide variety of codes to simulate an ever wider array of physical problems. The importance of radiation feedback in the problem of star formation cannot be overstated. It is an essential process in the formation regulation of star formation rates and efficiencies on galactic scales, the resulting effects on the structure and evolution of galaxies \citep{Agertz+2013}, the formation of star clusters and possible regulation of the IMF \citep{MacLowKlessen2004,McKeeOstriker2007}, the formation of massive stars \citep{Beuther+2007,Krumholz+2007a,Kuiper+2010a,Kuiper+2011,KuiperYorke2013a,KuiperYorke2013b}, and the heating and chemistry of protoplanetary disks and implications for planet formation \citep{AikawaHerbst1999,Fogel+2011}.

Radiation feedback's relevance for star formation has long been understood. The depletion time for a molecular cloud $t_{\mathrm{dep}} = M_{\mathrm{gas}}/\dot{M}$ is 1--3 orders of magnitude longer than the cloud's freefall time $t_{\mathrm{ff}} \sim 1/\sqrt{G\rho}$ \citep{KrumholzTan2007,Evans+2009}, yet simulations that do not include any feedback mechanisms routinely see unrealistically high star formation efficiencies, with $t_{\mathrm{dep}} \sim t_{\mathrm{ff}}$.

We focus here on radiation feedback, which has been repeatedly demonstrated in simulations as an effective means of suppressing star formation, even radiation from low-mass stars \citep{Offner+2009}. While magnetic fields have been suggested as a mechanism for slowing gravitational collapse, \citet{Crutcher2012} shows that many molecular clouds are 2--3 times supercritical to gravitational collapse. Magnetic fields have the effect of suppressing star formation \citep{TilleyPudritz2007,PriceBate2009}, and a careful treatment of them must be done in any accurate star formation simulation, but it is one of several critical processes in play. Other stellar feedback mechanisms have also been studied, including winds, outflows \citep{KuiperYorkeTurner2014,Peters+2014}, ionization \citep{Peters+2010a}, radiation pressure, and supernovae, but many of these are at least in part tied to the stellar radiation \citep{Murray+2010}.

We focus on the problem of thermal and momentum radiation feedback and the challenge of implementing highly accurate radiation hydrodynamics into a grid code for general simulations of star formation, leaving ionization effects to future work. Radiation hydrodynamics (RHD) methods have been implemented into several grid codes, e.g.\ {\sc zeus} \citep{TurnerStone2001}, {\sc athena} \citep{SkinnerOstriker2013}, {\sc orion} \citep{Krumholz+2007b}, {\sc ramses} \citep{Commercon+2011, Rosdahl+2013},  and {\sc pluto} \citep{Mignone2007} by \citet{Kuiper+2010b,Flock+2013,Kolb+2013}.

Most implementations of radiation hydrodynamics have been limited to the gray flux-limited diffusion (FLD) approximation, in which the radiation flux is proportional to the gradient of the radiation energy, $\grad E_r$. In this limit, it is usually assumed that the radiation and matter internal energies are tightly coupled, with the radiation and gas temperatures being equal. Further, the energy from stellar sources is deposited within some kernel centred on the radiation sources.

The problem with this approach is that the geometry of the environment surrounding stellar sources of radiation is rarely spherically symmetrical. Stars create outflow cavities and \HII regions that are optically thin to radiation. Meanwhile, protostars accrete material through a rotating disk structure that is usually highly optically thick. Outflow cavities provide an outlet for the radiation flux---the so-called ``flashlight effect'' described in \citet{YorkeSonnhalter2002}, \citet{Krumholz+2005b}, and \citet[submitted]{KuiperYorkeTurner2014}. Protostellar jets are on the scale of 100--1000 AU or even parsec scales for massive protostars, and \HII regions can easily approach the parsec scale as well.

Above all, the birth environments of stars contain supersonic turbulence, which gives rise to filaments that allow stellar radiation a path of easy egress into the optically thin cavities between filaments while remaining optically thick along the direction of the filament. Typical implementations of FLD schemes will fail to take into account this high degree of asymmetry when depositing stellar radiation energy. This may miss important effects. 

A particularly important problem in this regard is the formation of a massive star and its associated accretion disk. The intense radiation field from the forming star heats and ionizes the infalling envelope but does far less damage to the highly optically thick accretion disk through which most of gas, destined for the star, flows. Disk accretion therefore becomes the chief mechanism by which massive stars continue to accrete material despite their high luminosity \citep{Kuiper+2010a}. The accretion disk creates an environment with sharp transition regions between optically thick and optically thin, where traditional FLD codes are most inaccurate \citep{KuiperKlessen2013}. To address this high degree of asymmetry, \citet{Kuiper+2010b} implemented a radiation transfer scheme in {\sc pluto} that combined a 1D multifrequency raytrace with a gray FLD code. The method was implemented for spherical polar grids, which meant that the raytrace needed only be carried out in the radial direction. 

As an example application of this, \citet{Kuiper+2012} studied the stability of radiation-pressure-dominated cavities around massive protostars. Radiatively-driven outflows have the potential to remove a significant amount of mass from the stellar environment that would otherwise be accreted by the protostar. Several mechanisms have been proposed for how a massive protostar could accrete material beyond its Eddington limit. \citet{Krumholz2009} proposed a ``radiative Rayleigh-Taylor instability'' in which the radiatively-driven shell becomes unstable and material is able resume gravitational infall. But \citet{Kuiper+2012} argued that a gray FLD-only radiative transfer scheme underestimates the radiative forces acting on the shell.

We now generalize the method to 3D Cartesian grids with adaptive mesh refinement (AMR) and implement a hybrid raytrace/FLD method in the \FLASH astrophysics code. This method extends to multiple sources and does not rely on any special geometry, allowing for the treatment of more general problems such as star cluster formation. Moreover, for the FLD solver step, we implement a two-temperature (2T) radiation transport scheme; that is, we do not force the radiation temperature and matter temperature to be equal everywhere. We discuss the general theory of this approach in section \ref{sec:theory}; the equations to be solved and the numerical methods for \FLASH in section \ref{sec:numerical_methods}; the tests of our radiation transport scheme in sections \ref{sec:diffusion_tests}, \ref{sec:static_disk_test}, and \ref{sec:multiple_sources_tests}; and our final thoughts and discussion in section \ref{sec:conclusion}.

\section{Theory} \label{sec:theory}

The general idea to split the radiation field into a direct component and a diffuse or scattered component is an old one \citep[see, e.g.,][]{WolfireCassinelli1986,Murray+1994,EdgarClarke2003}. The direct component dominates in the optically thin regions, such as in the outflow cavities created by massive stars, or in \HII regions, where the temperature of the radiation field is that of the stellar photosphere. Within a few optical depths, inside the optically thick regions, the radiation field becomes dominated by the diffuse, thermal component of the radiation field. The main advantage of splitting the radiation field in this way is accuracy \citep{Murray+1994}. Raytracing is a solution to the radiative transfer problem, whereas FLD is a convenient approximation for the sake of computation. In the optically thin regions such as those noted above, a direct raytrace ignoring scattering is an excellent way of accurately calculating the radiation field, whereas the gray FLD is accurate inside optically thick regions where the temperature of the radiation is equilibrated to the matter temperature. In complex morphologies, with interspersed regions of varying optical depth, FLD is not at all accurate. 

This general splitting approach is equivalent to extracting the first absorption event from the FLD solver and allowing the radiation flux to be calculated by a raytracing scheme instead, with all re-emission and secondary absorption handled by the FLD solver. This improves the accuracy in precisely the region where the FLD approximation is worst, namely in regions directly irradiated by discrete radiation sources such as stars.

A hybrid radiation transport scheme splits the flux term,
\begin{equation}
\vec{F} = \vec{F}_* + \vec{F}_{\mathrm{th}},
\end{equation}
into a direct (stellar) component $\vec{F}_*$ and a thermal radiation component $\vec{F}_{\mathrm{th}}$. FLD implementations treat only the transport of a single radiation flux proportional to the gradient in the radiation energy $E_r$,
\begin{equation}
\vec{F}_{\mathrm{th}} \propto - \grad E_r.
\end{equation}
Hybrid schemes decompose the radiation field, transporting the direct component via a raytracer (see \ref{sec:irradiation}) and the indirect component via a diffusion equation. The radiation transport equation, integrated over all solid angle, with the operator-split hydrodynamic terms removed, is
\begin{equation}
\partial_t E_r + \grad \cdot \vec{F}_r = \kappa_P \rho \left(4 \pi B - c E_r\right),
\end{equation}
with $E_r$ is the radiation energy density, $\vec{F}_r$ the flux of the radiation energy, $\sigma_P$ the Planck opacity, $B = B(T)$ the Planck function, and $c$ the speed of light.

In a split scheme the direct component $\grad \cdot \vec{F}_*$ is calculated everywhere for all sources. We describe how this is done in section \ref{sec:irradiation}. How the thermal component is handled is the subject of section \ref{sec:FLD}.

In order to address problems of multiple star formation, as well as the formation of star clusters, a more general approach is needed. Specifically, we generalize the hybrid radiation transfer approach to Cartesian grids with AMR. These kinds of codes already have the ability to follow the gravitational collapse of multiple regions within a simulation, and have excellent implementations of turbulent and MHD processes. They also enable further study of  radiatively-driven shells and outflows with adaptive resolution.

The scope of the current paper is the implementation of the general hybrid radiation transfer scheme in an AMR grid code, and tests of its accuracy. As noted in the introduction, there are many important applications of our code which we will take up in subsequent papers, including the formation of a massive star in a turbulent, magnetized, collapsing medium, ionization feedback, and the formation of star clusters.   

\section{Numerical methodology} \label{sec:numerical_methods}
\subsection{FLASH} \label{sec:FLASH}

We use the publicly available \FLASH high-performance general application physics code, currently in its 4th major version \citep{Fryxell+2000,Dubey+2009}. The code is modular, and has physics capabilities that now include 2T radiation hydrodynamics (our contribution), magnetohydrodynamics, multi-group flux-limited diffusion, self-gravity, and a variety of options for the equation of state.

The code has been expanded to include sink particles, originally implemented in version 2.5 \citep{Federrath+2010}, and then later ported to version 4.0 \citep{Safranek-Shrader2012}. Multigroup flux-limited diffusion for radiation hydrodynamics was added in version 4.0 for the study of high-energy density physics (HEDP), such as radiative shocks and laser energy deposition experiments. Despite these original motivations, the code is general and can be applied to astrophysical scenarios.

A time-independent raytracer with hybrid characteristics was added in version 2.5 by \citet{Rijkhorst+2006} and then modified for the study of collapse calculations by \citet{Peters+2010a}. The raytracer was used to calculate photoelectric heating and photoionization rates, as well as heating by optically thin non-ionizing radiation, to study \HII regions and ionization feedback in molecular clouds \citep{Peters+2010a,Peters+2010b,Peters+2010c,Peters+2011,Peters+2012,Klassen+2012a}. However, this approach is not suitable to propagate non-ionizing radiation in optically thick clouds. 

We ported this raytracer into the present version of the \FLASH code in order to calculate the irradiation of gas and dust by point sources within the domain and solve the radiation hydrodynamic equations. 

\FLASH solves the fluid equations on an Eulerian mesh with adaptive mesh refinement (AMR) using the piecewise parabolic method \citep{ColellaWoodward1984}. The method is well-suited for dealing with shocks. The diffusion equation is solved implicitly via the generalized minimal residual (GMRES) method. The refinement criterion used by \FLASH is an error estimator based on \citet{Lohner1987}, which is a modified second derivative normalized by the average of the gradient over one computational cell. It has the advantage of being dimensionless and local, so one has the flexibility to choose the grid variable upon which to refine.

In this paper we describe the implementation of a hybrid radiation transfer scheme, involving raytracing coupled to a flux-limited diffusion (FLD) solver. The raytracer finds the flux each cell receives from all point sources (e.g. stars) in the simulation while the FLD solver evolves the diffuse radiation field. The combined radiation field is used to update the matter temperature.

In the sections below we recapitulate the basic theory and how these equations are implemented in \FLASH.

\subsection{Radiation hydrodynamics}\label{sec:radiation_hydro}

If the radiation fields are assumed to have a blackbody spectrum, then in the absence of magnetic and gravitational fields, and assuming local thermodynamic equilibrium (LTE), the frequency- and angle-integrated radiation hydrodynamic equations in the comoving frame are \citep{TurnerStone2001,MihalasMihalas84} 
%
% THE EQUATIONS OF RADIATION HYDRODYNAMICS
%
\begin{eqnarray}
% Conservation of Mass:
\label{eqn:mass_conservation}
&\frac{D \rho}{Dt} + \rho \div \vec{v} = 0 \\
% Conservation of Momentum:
\label{eqn:momentum_conservation}
&\rho \frac{D \vec{v}}{Dt} = - \grad p + \frac{1}{c}\kappa_P \rho \vec{F}_r \\
% Internal energy transport:
\label{eqn:energy_transport}
&\rho \frac{D}{Dt}\left(\frac{e}{\rho}\right) + p \div \vec{v} = - \kappa_P \rho (4 \pi B - c E_r)\\
% Radiation energy transport:
\label{eqn:radiation_energy_transport}
&\rho \frac{D}{Dt}\left(\frac{E_r}{\rho}\right) + \div \vec{F}_r + \grad \vec{v} \vec{:} \vec{P} = \kappa_P \rho(4 \pi B - c E_r)% \\
% Flux transport: (Leaving out flux transport, since our method does not solve this equation; Eddington approximation)
%\label{eqn:flux_transport}
%&\frac{\rho}{c^2}\frac{D}{Dt}\left(\frac{\vec{F}_r}{\rho}\right) = - \div \vec{P} - \frac{1}{c}\kappa_P \rho \vec{F}_r
\end{eqnarray}
where $D/Dt \equiv \partial/\partial t + \vec{v \cdot \nabla}$ is the convective derivative. The fluid variables $\rho, e, \vec{v}$, and $p$ are the matter density, internal energy density, fluid velocity, and scalar pressure, respectively, while $E_r, \vec{F}_r$, and $\vec{P}$ are the total frequency-integrated radiation energy, flux, and pressure, respectively. Equations \ref{eqn:mass_conservation} and \ref{eqn:momentum_conservation} express the conservation of mass and momentum, respectively, where the radiation applies a force proportional to $\kappa_P \rho \vec{F}_r/c$, where $\kappa_P$ is the Planck mean opacitiy defined in equation \ref{eqn:planck_mean} below.

The hydrodynamic equations must also be closed via an equation of state. We use a gamma-law equation of state for simple ideal gases. The gas pressure $P$, density $\rho$, internal energy $\epsilon$, and gas temperature $T$ are related by the equations,
\begin{equation}\label{eqn:eos}
P = (\gamma - 1)\rho \epsilon = \frac{N_a k_B}{\mu}\rho T,
\end{equation}
with adiabatic index $\gamma = 5/3$. $N_a$ is Avogadro's number, $k_B$ is the Boltzmann constant, and $\mu$ is the mean molecular weight.

The coupling between internal and radiation energy is expressed by equations \ref{eqn:energy_transport} and \ref{eqn:radiation_energy_transport}. Coupling comes through the emission and absorption of radiation energy. The assumption of LTE lets us write the source function as the Planck function, $B$, so that matter emits thermal radiation proportional to the rate,
\begin{equation*}
4 \pi \kappa_P \rho B(T) \equiv \kappa_P \rho c a T^4,
\end{equation*}
where $\kappa_P$ is the Planck mean opacity,
\begin{equation}\label{eqn:planck_mean}
\kappa_P = \frac{\int_0^\infty d \nu \kappa_{\nu}B_{\nu}(T)}{B(T)},
\end{equation}
with
\begin{equation}
B_{\nu}(T) = \frac{2 h \nu^3 / c^2}{e^{h\nu/k_B T} - 1}.
\end{equation}

Meanwhile, matter is absorbing radiation out of the thermal field at a rate $\kappa_P \rho c E_r$. Energy lost by the radiation field is gained by the matter and vice versa. We will use the flux-limited diffusion approximation to relate the radiation flux $\vec{F}_r$ to the radiation energy $E_r$.

In practice, many grid codes implement the flux-limited diffusion approximation \citep{TurnerStone2001,Krumholz+2007b,Commercon+2011}, which relates the radiation flux $\vec{F}_r$ to the radiation energy $E_r$. %Under the Eddington approximation, the radiation flux is
%\begin{equation}
%\vec{F} = - \frac{c}{3\chi}\grad{E}_r,
%\end{equation}
%where $\chi$ is the total opacity. This expression suffers from the problem that in optically thin regions $(\chi \rightarrow 0)$ the flux becomes infinite. Information about the radiation field cannot travel faster than the speed of light. The Eddington approximation as stated becomes invalid and we require a new expression for the radiation flux. In order to retain most of the simplicity of this approach without also violating causality \citet{LevermorePomraning1981} introduced a flux limiter, a dimensionless scalar function of the radiation energy to smoothly bridge the optically thin and optically thick regimes while ensuring that energy propagation never exceeded the speed of light. This approach is described in section \ref{sec:FLD}.
%
In the FLD approximation, the radiation and matter components may be thought of as two fluids, each with their own equation of state, internal energy, temperature, and pressure. The matter component may be further split into ion and electron components, which exchange energy via collisions. This is three-temperature, or ``3T'' approach is used in the modeling of laboratory plasmas. The \FLASH code was extended in version 4 to include 3T radiation hydrodynamics \citep{Lamb+2010,Fatenejad+2012,Kumar+2011}. Energy exchange under the 3T model is included for reference in Appendix \ref{sec:appendix}. In astrophysics, we are more concerned with modeling the gas/dust mixture of the ISM, with the gas sometimes ionized. By setting a unified matter temperature in \FLASH, with $T_{\mathrm{ion}} = T_{\mathrm{ele}} \equiv T_{\mathrm{gas/dust}}$, we return to a 2T model, with only matter and radiation temperatures varying and exchanging energy. 

Radiation only behaves like a fluid when it is tightly coupled to matter, which occurs only in very optically thick regions. Hybrid radiation schemes more accurately transport the radiation energy through the computational domain.

\subsection{Flux-limited diffusion} \label{sec:FLD}

Under the flux-limited diffusion approximation \citep{LevermorePomraning1981,Bodenheimer+1990}, the flux of radiation is proportional to the gradient in the radiation energy density, 
\begin{equation}\label{eqn:fld_approximation}
\vec{F}_r = - D \grad E_r, 
\end{equation}
where
\begin{equation}
D = \frac{\lambda c}{\kappa_R \rho}
\end{equation} 
is the diffusion coefficient. $\kappa_R$ is the Rosseland mean opacity and $\lambda$ is the flux limiter that bridges the radiation diffusion rate between the optically thin and optically thick regimes. In the extreme optically thick limit, $\lambda \rightarrow 1/3$, which is the diffusion limit. In the extreme optically thin limit, the flux of radiation becomes $F_r = cE_r$, the free-streaming limit.

We use the \citet{LevermorePomraning1981} flux limiter, one of the most commonly used,
\begin{equation}
\lambda = \frac{2+R}{6+3R+R^2},
\end{equation}
with
\begin{equation}
R = \frac{|\grad E_r|}{\kappa_R \rho E_r}.
\end{equation}
Other flux limiter are possible, such as \citet{Minerbo1978}, another popular choice. Different flux limiters result from different assumptions about the angular distribution of the specific intensity \citep{TurnerStone2001}.

\FLASH solves the radiation diffusion equation \ref{eqn:radiation_energy_transport} using a general implicit diffusion solver. Using the method of operator splitting, terms such as advection and hydrodynamic work are handled by the code's hydro solver. The radiation diffusion solver then updates the radiation energy equation by solving
\begin{equation}\label{eqn:radiation_diffusion}
\frac{\partial E_r}{\partial t} + \div \left(\frac{\lambda c}{\kappa_R \rho} \grad E_r\right) = \kappa_P \rho c \left(a T^4 - E_r \right)
\end{equation}
where $T$ is the gas temperature. 

To solve this equation, we use the method of the generalized minimum residual (GMRES) by \citet{GMRES}. It is included in \FLASH via the \HYPRE library \citep{HYPRE}. The GMRES method is the same one employed by \citet{Kuiper+2010b} in their hybrid radiation-transport scheme. It belongs to the class of Krylov subspace methods for iteratively solving systems of linear equations of the form $A \vec{x} = \vec{b}$, where the matrix $A$ is large and must be inverted. Matrix inversion is, in general, a very computationally expensive operation, but because $A$ is also a sparse matrix, methods exist for rapidly computing an approximate inverse with relatively high accuracy, outperforming methods such as conjugate gradient (CG) and successive over-relaxation (SOR). The \HYPRE library that \FLASH uses is a collection of sparse matrix solvers for massively parallel computers. 

A difference in matter and radiation temperatures in equation \ref{eqn:radiation_diffusion} results in an energy ``excess'' on the right-hand side. We must update the matter temperature due to the action of the combined radiation fields. Restating equations \ref{eqn:energy_transport} and \ref{eqn:radiation_energy_transport}, with operator-split hydrodynamic terms removed, and a stellar radiation source term added, we have

\begin{align}
\label{eqn:eint_evol}
 \partial_t \rho \epsilon &= - \kappa_P \rho c \left(a_R T^4 - E_r \right) - \div \vec{F}_* \\
\label{eqn:erad_evol}
 \partial_t E_r + \grad \cdot \vec{F}_r &= + \kappa_P \rho c \left(a_R T^4 - E_r \right)
\end{align}

Discretizing equation \ref{eqn:eint_evol} and \ref{eqn:erad_evol}, we have
\begin{equation}\label{eqn:discretized_eint}
\tfrac{\rho c_V T^{n+1} - \rho c_V T^{n}}{\Delta t} = -\kappa_P^n \rho c \left( a_R \left(T^{n+1}\right)^4 - E_r^{n+1}\right) - \div \vec{F}_*,
\end{equation}
and
\begin{equation}\label{eqn:discretized_erad}
\tfrac{E_r^{n+1} - E_r^{n}}{\Delta t} - \nabla \cdot \left( D^{n} \nabla E_r^{n+1}\right) = + \kappa \rho c \left( a_R \left(T^{n+1}\right)^4 - E_r^{n+1}\right)
\end{equation}
recalling that the specific internal energy $\epsilon = c_V T$, with $c_V$ being the specific heat capacity of the matter. Variables with superscript indices $n$ and $n+1$ take their values from before and after the implicit update, respectively. It is important to remember that the opacity $\kappa_P$ is temperature-dependent. The source term for stellar radiation takes the form $\div \vec{F}_*$ and represents the amount of stellar radiation energy absorbed at a given location in the grid from all sources.

The presence of the nonlinear term $(T^{n+1})^4$ makes it difficult to solve for the temperature in \ref{eqn:discretized_eint}. Assuming that the change in temperature is small, \citet{Commercon+2011} linearizes this term,
\begin{align}\label{eqn:linear_temp}
\left(T^{n+1}\right)^4 &= (T^n)^4\left(1+\frac{T^{n+1} - T^n}{T^n}\right)^4 \nonumber \\
                                 &\approx 4\left(T^n\right)^3 T^{n+1} - 3\left(T^n\right)^4.
\end{align}

This allows for equation \ref{eqn:linear_temp} to be substituted into equation \ref{eqn:discretized_eint} and for us to write an expression for the updated temperature,
\begin{small}
\begin{equation}\label{eqn:updated_temp}
T^{n+1} = \frac{3 a_R \alpha \left(T^n\right)^4 + \rho c_V T^n + \alpha E_r^{n+1} - \Delta t \div \vec{F}_*}{\rho c_V + 4 a_R \alpha \left(T^n\right)^3},
\end{equation}
\end{small}
where we have used $\alpha \equiv \kappa_P^n \rho c \Delta t$.

The irradiation term, $\div \vec{F}_*$, is supplied by the raytracer. Its calculation is the subject of section \ref{sec:irradiation}.

The raytrace is performed first in order to calculate $\div \vec{F}_*$. By substituting equation \ref{eqn:updated_temp} into \ref{eqn:discretized_erad} via the approximation of equation \ref{eqn:linear_temp}, and making appropriate simplifications, the FLD equation becomes
%\begin{align}\label{eqn:updated_FLD}
\begin{small}
\begin{multline}\label{eqn:updated_FLD}
\frac{E_r^{n+1} - E_r^{n}}{\Delta t} - \nabla \cdot \left( D^{n} \nabla E_r^{n+1}\right) + \\ \kappa \rho c \left[\frac{ \rho c_V T^{n}}{\rho c_V T^{n} + 4 \alpha a_R \left(T^{n}\right)^4}\right] E_r^{n+1} \\
= \kappa \rho c a_R \left(T^{n}\right)^4 \left[\frac{\rho c_V T^{n} - 4 \Delta t \nabla \cdot F_*}{\rho c_V T^{n} + 4 \alpha a_R \left(T^{n}\right)^4} \right],
\end{multline}
\end{small}
where we have gathered implicit terms on the left-hand side and explicit terms on the right-hand side.

We solve equation \ref{eqn:updated_FLD} via the diffusion solver to calculate the updated radiation energy density and temperature, then update the gas/dust temperature via \ref{eqn:updated_temp}. This completes the radiation transfer update for the current timestep.

\subsection{Stellar irradiation} \label{sec:irradiation}

When the effects of scattering and emission are ignored, the equation for the specific intensity of radiation from a point source along a ray assumes a very simple form,
\begin{equation}
I(r) = I_0 e^{-\tau(r)},
\end{equation}
where $I_0$ is the specific intensity of the source and $\tau(r)$ is the optical depth from the source up to $r$:
\begin{equation}
\tau(r) = \int_{R_*}^r \kappa_P(T_*) \rho(r^{\prime}) \, \mathrm{d}r^{\prime}
\end{equation} 
The opacity $\kappa_P$ is a function of temperature and the optical depth is calculated using the temperature of the radiation source, e.g. the effective temperature of a star. $\rho$ is the usual gas density. Rays are traced from individual cells in the computational grid back to the point sources (usually stars) via a characteristics-based method described in \citet{Rijkhorst+2006}.

In characteristics-based raytracing, there are two approaches to computing the column density toward sources, each with its own advantages and disadvantages. \textit{Long characteristics} involves tracing rays from the source to each cell in the computational domain. This method is very accurate, but involves many redundant calculations of the column density through cells intersected by similar rays near the source. \textit{Short characteristics} tries to mitigate the redundant calculations by tracing rays only across the length of each grid cell in the direction of the source, then interpolating upwind toward the source to calculate the total column density. The disadvantage of this approach is that the upwind values need to be known ahead of time, which imposes an order on the calculations and makes scaling the code to many processors very problematic. Some amount of numerical diffusivity is also introduced in this approach on account of interpolating column densities at each cell. 

As a compromise between these two methods, \citet{Rijkhorst+2006} developed a method they called \textit{hybrid characteristics} that minimizes the number of redundant calculations and can be parallelized. The local contributions to the column density are computed for each block of $n\times n\times n$ cells. Then the values on the block faces are computed and communicated to all processors. Finally, the total column density at each cell is computed by adding the local and interpolated face values.

\citet{Peters+2010a} have improved the method to allow collapse simulations with arbitrary many refinement levels and sources of radiation. They added the propagation of (optically thin) non-ionizing radiation and coupled the respective heating terms to a prescription of molecular and dust cooling \citep{Banerjee+2006}. Furthermore, they have linked the radiation module to sink particles \citep{Federrath+2010} as sources of radiation and implemented a simple prestellar model to determine the value of the stellar and accretion luminosities. In our current implementation, we no longer use a separate cooling function because the thermal evolution is already implicit in the coupled equations of internal and radiation energy, as described in section \ref{sec:FLD}.

The stellar flux a distance $r$ is given by
\begin{equation}\label{eqn:stellar_flux}
F_*(r) = F_*(R_*) \left(\frac{R_*}{r}\right)^2 \exp\left(-\tau(r)\right),
\end{equation}
where $R_*$ is the stellar radius, and $F_*(R_*)$ is the flux at the stellar surface, given by
\begin{equation}
F_*(R_*) = \sigma T_*^4.
\end{equation}
$T_*$ is the effective temperature of the stellar surface and $\sigma$ is the Stefan-Boltzmann constant.

At this time, our raytracing is purely gray; we use the \citet{DraineLee1984} dust model and compute the temperature-dependent frequency-averaged opacities, assuming a 1\% dust-to-gas ratio. Multifrequency raytracing is relatively straightforward to implement, but the computational costs increase linearly with the number of frequency bins. We leave the implementation of multifrequency effects to a future paper.

Using the stellar flux computed at each cell, we estimate an ``irradiation'' term, i.e. the amount of stellar flux absorbed by a given cell,
\begin{equation}
\grad \cdot F_*(r) \approx - \frac{\left(1 - e^{\tau_{\mathrm{local}}}\right)}{\Delta r} F_*(r),
\end{equation}
where $\tau_{\mathrm{local}}$ is the ``local'' contribution to the optical depth through the cell, i.e.
\begin{equation}
\tau_{\mathrm{local}} \approx \kappa_P(T_*) \rho \, \Delta r,
\end{equation}
and $\Delta r$ is the distance traced by the ray through a grid cell. For numerical stability, when $\tau_{\mathrm{local}}$ is very small, e.g. $10^{-5}$, we estimate the irradiation by Taylor-expanding the exponential,
\begin{equation}\label{eqn:irradiation}
\grad \cdot F_*(r) \approx - \kappa_P \rho F_*(r),
\end{equation}
which is the optically thin limit. 

\subsection{Opacities} \label{sec:opacities}

The primary source of opacity in the interstellar medium is dust grains. We use the tabulated optical properties of graphite and silicate dust grains from \citet{DraineLee1984}. They evaluate absorption cross-sections for particles with sizes between 0.003 and 1.0 microns at wavelengths between 300 \AA~and 1000 microns. These tabulated dust properties are widely used both in simulations \citep[e.g.][]{Pascucci+2004,Offner+2009,Flock+2013} and in observational work \citep[e.g.][]{Nielbock+2012}.

\begin{figure}
\includegraphics[width=88mm]{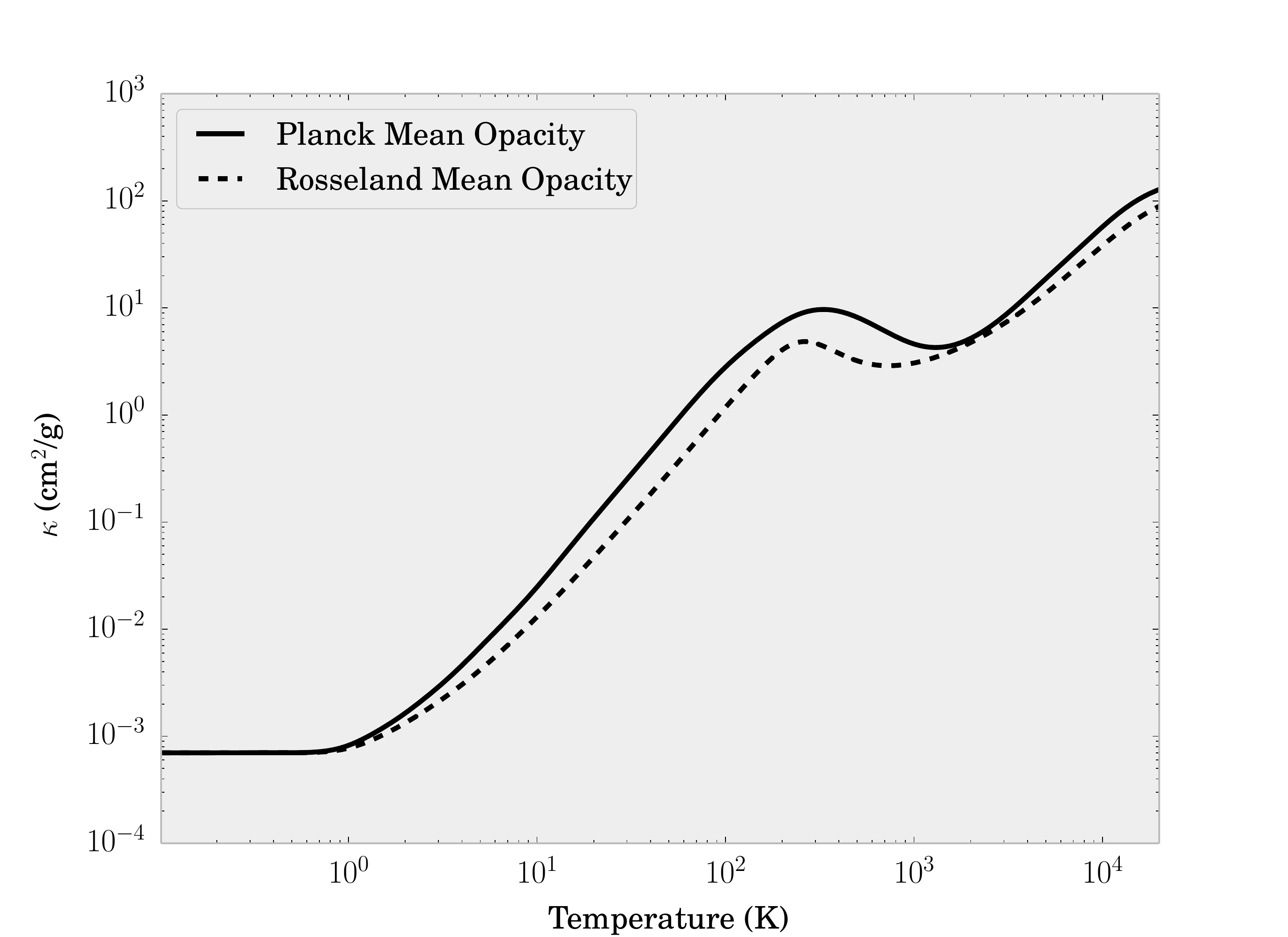}
\caption{Frequency-averaged opacity (in units of cm$^2$ per gram gas) as a function of temperature for a 1\% mixture of interstellar dust grains with gas, based on the model by \citet{DraineLee1984}.}
\label{fig:opacity_tables}
\end{figure}

We assume a dust to gas ratio of 1\% and integrate the dust opacity over a wavelength range of 1 \AA--1000 microns. This is done as a function of temperature, resulting in ``gray'' (frequency-averaged) temperature-dependent Planck and Rosseland mean opacity tables. Figure \ref{fig:opacity_tables} shows, e.g., the opacity $\kappa$ over a range of temperatures from 0.1 K to 2000 K. We see that these opacities cover many orders of magnitude.

\subsection{Radiation pressure} \label{sec:radiation_pressure}

Both the direct radiation field and the diffuse radiation field contribute to the radiation pressure. In the case of the diffuse (thermal) radiation field, \FLASH makes the Eddington approximation, 
\begin{equation}\label{eqn:eddington_approximation}
P_{\mathrm{rad}} = \frac{1}{3} E_r = \frac{aT_{\mathrm{rad}}^4}{3},
\end{equation}
where $E_r$ is the radiation energy density and $T_{\mathrm{rad}}$ is the corresponding temperature. Additional momentum is added to the gas from the direct component of the radiation field. We take the stellar fluxes computed by the raytracer (equation \ref{eqn:stellar_flux}). These exert a body force
\begin{equation}\label{eqn:radiation_force_density}
f_{\mathrm{rad}} = \rho \kappa_P(T_*) \frac{F_*}{c} = - \frac{\div \vec{F}_*}{c},
\end{equation}
where $f_{\mathrm{rad}}$ is the force density and $F_*$ is the direct stellar radiation flux. The temperature of the direct radiation field is that of the source, $T_* = T_{\mathrm{eff}}$, the effective stellar surface temperature.

From equation \ref{eqn:radiation_force_density} we see that the radiation force density is proportional to the absorbed radiation energy computed by the raytracer, i.e.~the force felt by the gas and dust in a local region is proportional to its absorbed stellar radiation energy. 

We do not expect our radiation force to change much if it were to be computed with a multifrequency raytracer. The radiation force is proportional to the absorbed flux. In the multifrequency case, infrared radiation can penetrate further into the gas, but also carries less energy. Most of the higher frequencies are all absorbed in the same (few) grid cells. Tests by \citet{Kuiper+2010b} also showed very little deviation between the gray and multifrequency cases.

\subsection{Summary of the hybrid radiation hydrodynamics method}

In summary, we treat radiation hydrodynamics by complementing the fluid dynamics equations to include the effects of radiation (equations \ref{eqn:mass_conservation} through \ref{eqn:radiation_energy_transport}. We have decomposed the radiation field into a direct stellar component and an indirect diffuse component. The method of raytracing is used to calculate the direct stellar radiation field. The method of flux-limited diffusion is used to transport radiation in the diffuse radiation field.

Matter and radiation are coupled by emission and absorption processes, while stars represent sources of radiation energy. This relationship is given in equations \ref{eqn:eint_evol} and \ref{eqn:erad_evol}.

We have discretized and linearized equation \ref{eqn:eint_evol} and \ref{eqn:erad_evol} to derive an expression for evolving the matter temperature, given absorption and emission of radiation, the presence of a thermal radiation field, and the flux from discrete stellar sources. This is expressed in equations \ref{eqn:updated_temp} and \ref{eqn:updated_FLD}.

The next sections describe tests of the radiation transport routines and the reliability of our method.

\section{Tests of thermal radiation diffusion} \label{sec:diffusion_tests}

\subsection{Thermal radiative equilibration} \label{sec:radiation_equilibrium}

To test the accuracy of the matter/radiation coupling, we set up a unit simulation volume with a uniform gas density of $\rho = 10^{-7}$ g/cm$^3$ initially out of equilibrium with the radiation field \citep{TurnerStone2001}. If the radiation energy dominates the total energy, then any radiation energy absorbed or emitted is relatively small and the radiation field can be said to be unchanging. The matter energy evolves according to equation \ref{eqn:eint_evol}, but without the source term, i.e.

\begin{equation}\label{eqn:matter-radiation_coupling}
\frac{\partial e}{\partial t} = \chi c \left(E_r - a_R T^4 \right),
\end{equation}
where $e = \rho \epsilon$ is the volumetric matter energy density, and $\chi = \kappa_P \rho = 4 \times 10^{-8}$ cm$^{-1}$ is the absorption coefficient.

\begin{figure}
\includegraphics[width=88mm]{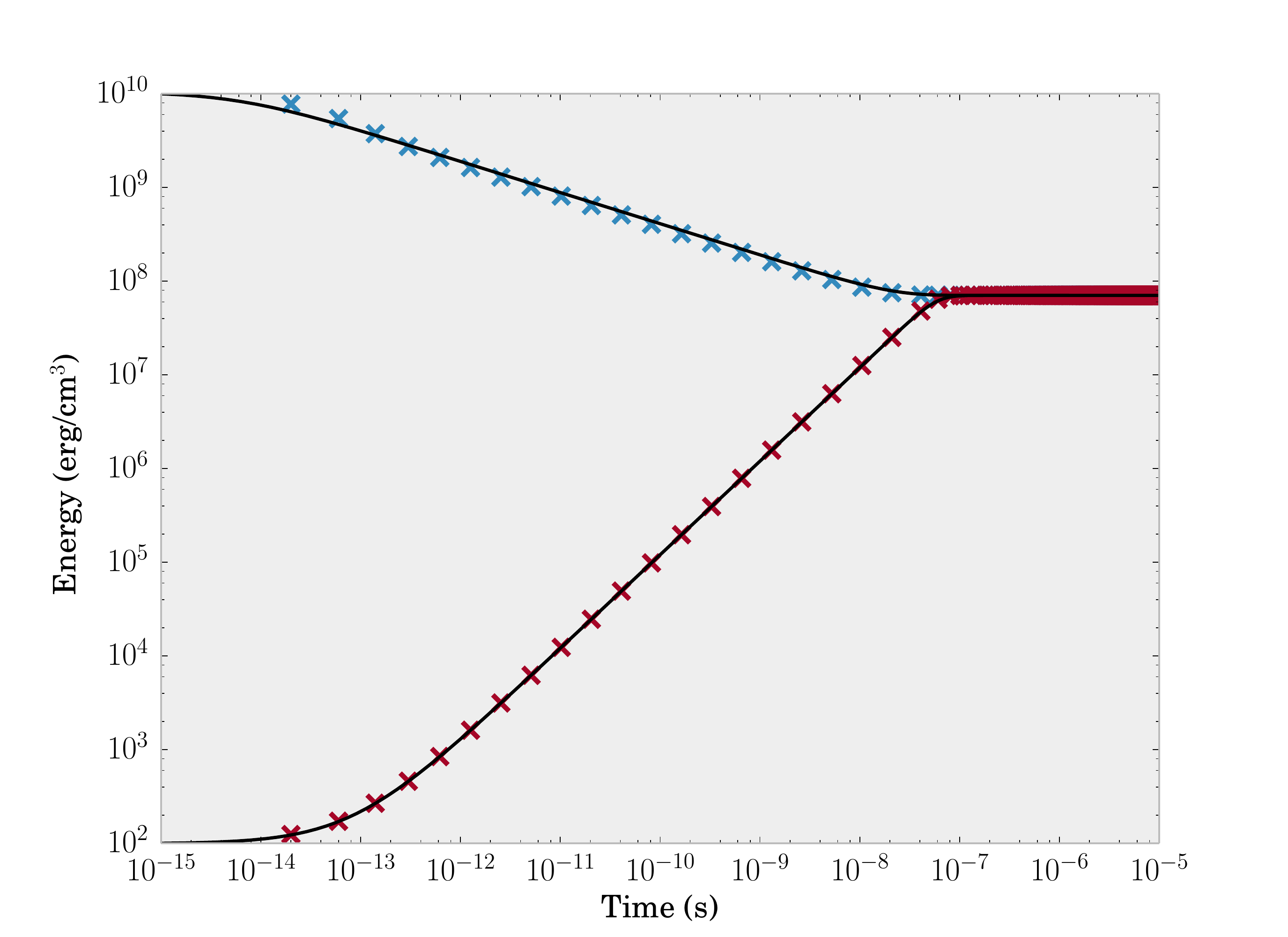}
\caption{Results from matter-radiation coupling tests. The initial radiation energy density is $E_r = 10^{12}$ ergs/cm$^3$. The matter internal energy density is initially out of thermal equilibrium with the radiation field. Crosses indicate simulation values at every time step, while the solid line is the analytical solution, assuming a constant $E_r$. Initial matter energy densities are $e = 10^{10}$ ergs/cm$^3$ (upper set) and $e = 10^2$ ergs/cm$^3$ (lower set).}
\label{fig:matter-radiation_coupling}
\end{figure}

We set the radiation energy density $E_r = 10^{12}$ ergs/cm$^3$, as in \citet{TurnerStone2001}. We solve equation \ref{eqn:matter-radiation_coupling} assuming a constant $E_r$ and plot the results in figure \ref{fig:matter-radiation_coupling} as the solid black lines. The initial matter energy density is $e = 10^{10}$ ergs/cm$^3$ in the case where the gas cools to equilibrium. In the gas warming case, the initial matter energy density is $e = 10^{2}$ ergs/cm$^3$.

The gas is assumed to be an ideal gas with $\gamma = 5/3$ and a mean mass per particle of $\mu = 0.6$.

Figure \ref{fig:matter-radiation_coupling} shows the results of numerical simulations with FLASH. The simulation has an adaptive timestep initially set to $\Delta t = 10^{-14}$ s. The maximum timestep size for this simulation was set to $\Delta t = 10^{-8}$ s. The total energy in the simulation volume is conserved to better than 1\% over the duration of the simulation.

\subsection{1D radiative shock tests} \label{sec:radiation_shock}

The treatment of radiative shocks is an important benchmark in many radiative transfer codes \citep{HayesNorman2003,WhitehouseBate2006,Gonzalez+2007,Kuiper+2010b,Commercon+2011}. We follow the setup described in \citet{Ensman1994} of a streaming fluid impinging on a wall, represented by a reflective boundary condition. The fluid is compressed and a shock wave travels in the upstream direction. The hot fluid radiates thermally, and the radiation field preheats the incoming fluid. By varying the speed of the incoming fluid, sub- or supercritical shocks can be formed. Criticality occurs when there is sufficient upstream radiation flux that the preshock temperature is equal to the postshock temperature. The fluid velocity at which this occurs is called the critical velocity. Numerical simulations can be compared to analytic arguments by \citet{MihalasMihalas84} for the gas temperature in various parts of the shock to check how well these shock features are being reproduced by the code.

The initial conditions are as follows: an ideal fluid ($\gamma = 5/3$) has a uniform mass density of $\rho_0 = 7.78 \times 10^{-10}$ g/cm$^3$, a mean molecular weight $\mu = 1$, and is at a uniform temperature of $T_0 = 10$ K. The domain size is $L = 7 \times 10^{10}$ cm.

For the subcritical shock, the fluid moves to the left with a speed $v = 6$ km/s. For the supercritical shock, $v = 20$ km/s. The fluid is given a uniform absorption coefficient of $\sigma = 3.1 \times 10^{-10}$ cm$^{-1}$. 

\begin{figure}
\includegraphics[width=88mm]{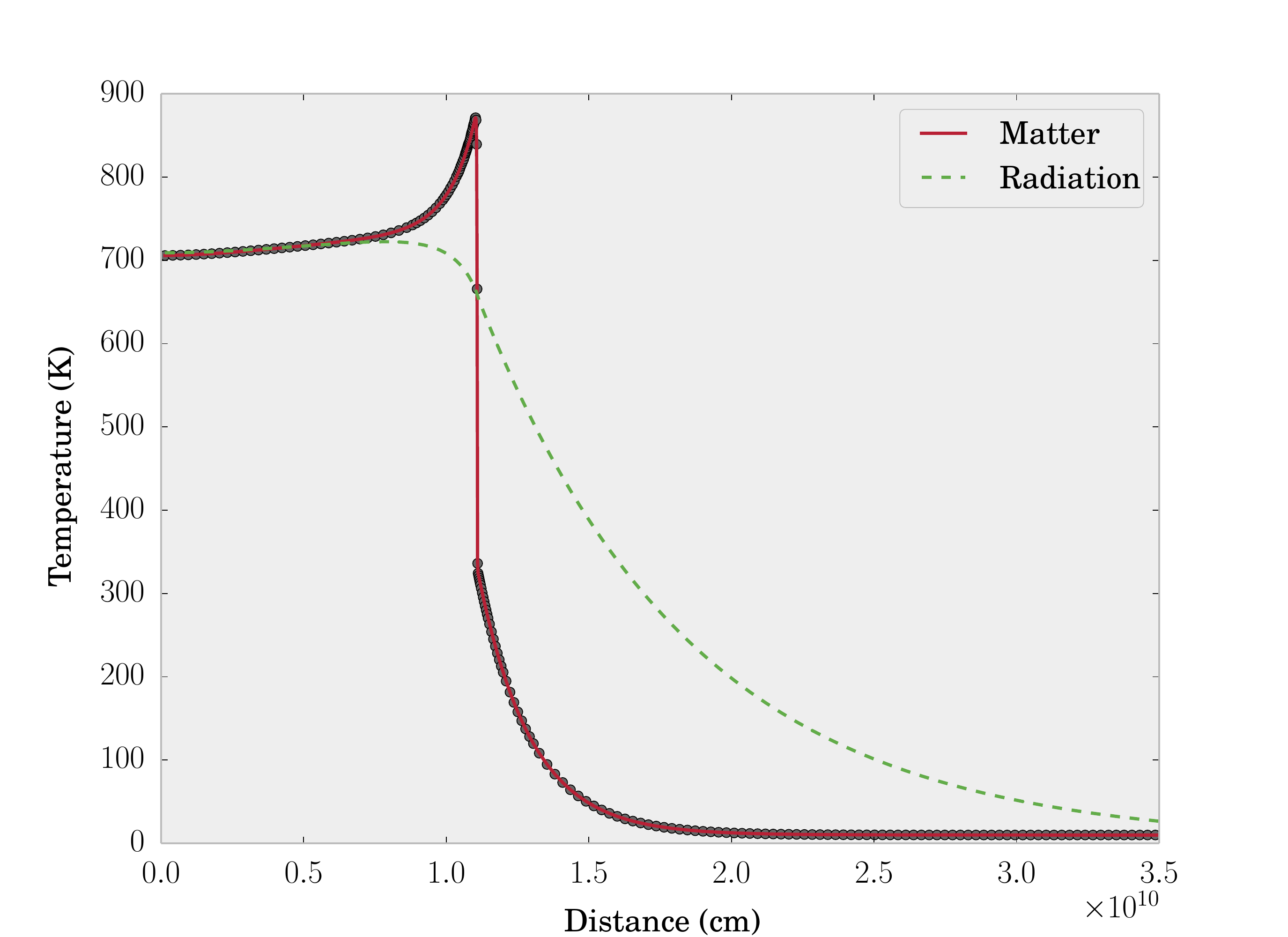}
\caption{Temperature profile for a subcritical radiative shock with a gas velocity of $v = 6$ km/s at time $t = 5.80 \times 10^{4}$ s. Red and green lines indicates matter and radiation temperatures, respectively. Circles along matter temperature indicate the grid resolution.}
\label{fig:subcritical_shock}
\end{figure}

\begin{figure}
\includegraphics[width=88mm]{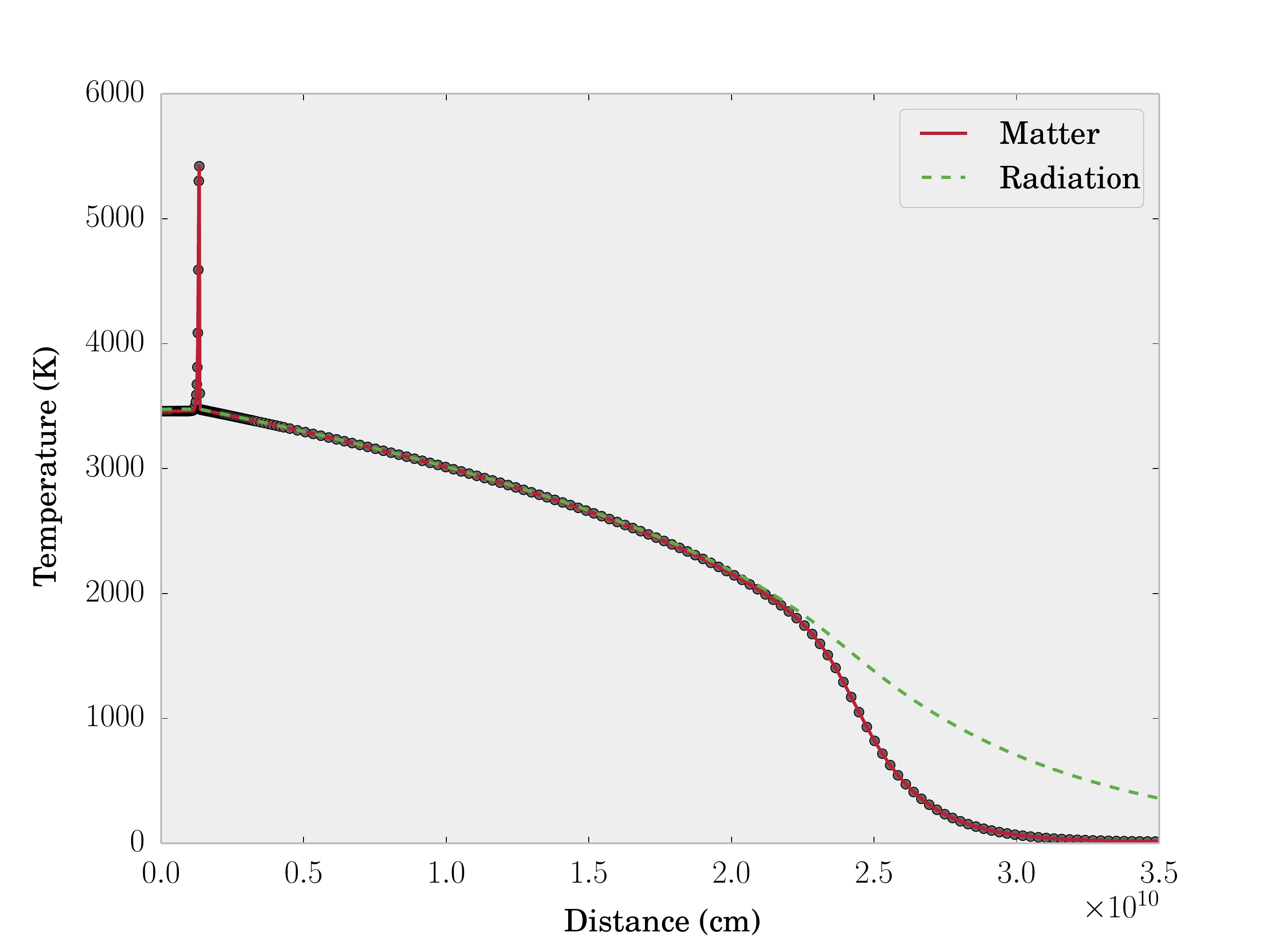}
\caption{Temperature profile for a supercritical radiative shock with a gas velocity of $v = 20$ km/s at time $t = 5.08 \times 10^{3}$ s. Lines and circles represent the same as in previous figure.}
\label{fig:supercritical_shock}
\end{figure}

Upon compression of the gas at the leftmost boundary, the postshock temperature $T_2$ increases and a radiative flux of order $\sigma_B T_2^4$ is produced, where $\sigma_B$ is the Stefan-Boltzmann constant. This radiation penetrates upstream to preheat the gas ahead of the shock front to a temperature $T_-$. This preheating can be clearly seen in Figure \ref{fig:subcritical_shock}. The shock is considered subcritical so long as $T_- < T_2$.

If the speed of the incoming gas $v$ is increased, there is greater preheating and the preshock temperature approaches the postshock temperature, $T_- \sim T_2$. These temperatures are equal for a critical shock. If the incoming gas speed is increased still further, the temperature of the preshock gas remains steady, while the radiative precursor is extended. This is a supercritical shock, as seen in Figure \ref{fig:supercritical_shock}.

For the subcritical case, if $T_- \ll T_2$, an approximate solution is given in \citet{MihalasMihalas84} for the postshock temperature,

\begin{equation}
T_2 \approx \frac{2\left(\gamma - 1\right)v^2}{R\left(\gamma + 1\right)^2},
\end{equation}
where $R = k_B / \mu m_H$ is the ideal gas constant.

For our subcritical case, with an incoming gas velocity of $v = 6$ km/s, this gives $T_2 \approx 812$ K. In our test case, the postshock temperature at the far-left of the domain is 706 K, or about 13\% less than the approximate value.

The preshock temperature can be approximated \citep{MihalasMihalas84} by
\begin{equation}
T_- \approx \frac{\gamma - 1}{\rho v R} \frac{2 \sigma_B T_2^4}{\sqrt{3}} \sim 279 \mbox{K.}
\end{equation}
The temperature spike at the subcritical shock front is approximated by
\begin{equation}
T_+ \approx T_2 + \frac{3 - \gamma}{\gamma + 1}T_- \sim 874 \mbox{K.}
\end{equation}

Our preshock temperature reaches 336 K, which is about 20\% warmer than the analytic approximation. Similar calculations in \citet{Kuiper+2010b} did not reproduce any preheating due to the 1-T approach in FLD. Our numerical shock temperature $T_+ \approx 871$ K agrees to within better than 1\%.

In the supercritical case, the radiation spike has collapsed to a thickness of less than about a photon mean free path. The temperature of the spike is $T_+ \approx 5420$ K, whereas the analytic approximation \citep{MihalasMihalas84} gives
\begin{equation}
T_+ \approx (3 - \gamma)T_2 \sim 4612 \mbox{K,}
\end{equation}
with which our numerical calculations agree to within 18\%. Comparing to other FLD implementations on AMR grids, \citet{Commercon+2011} matches the analytic estimates of the subcritical postshock temperature a little closer (to within 2\%). Our scheme captures the preshock heating that the 1-T scheme described in \citet{Kuiper+2010b} could not (\MAKEMAKE has since been made into a 2-T scheme), but not as closely as \citet{Commercon+2011}, which captures it to within 1\%. The subcritical shock temperature we capture comparably well to \citet{Commercon+2011}. While \FLASH has a different refinement criterion from the one used in \citet{Commercon+2011}, this is not likely the cause of the differences. More likely is the different choice of flux limiter (\citet{LevermorePomraning1981} vs \citet{Minerbo1978}), which, because of different assumptions about the angular dependence of the radiation field in a particular problem, can yield slightly different solutions, with the greatest differences seen in regions of intermediate to low optical depth \citep{TurnerStone2001}. It is not obvious which flux limiter is best for a given problem, but we have opted for one shared by \citet{Kuiper+2010b}.

%\section{Irradiation of a uniform density medium} \label{sec:uniform_medium_irradiation}
%
%To test whether our code is producing the correct matter temperatures, we consider a uniformly dense, homogeneous medium irradiated by a single stellar-type source located at one corner of the simulation volume.
%
%In the case of an optically thin medium, the luminosity $L = 4 \pi r^2 \sigma_{\mathrm{SB}} T^4$ is constant with radial distance $r$ from the stellar radiation source. The radiation flux,
%\begin{equation}
%F_r = \sigma_{\mathrm{SB}} T^4 = c E_r \propto r^{-2}.
%\end{equation}
%
%Therefore, since $E_r \propto T_{\mathrm{rad}}^4$, it holds that $T_{\mathrm{rad}} \propto r^{-1/2}$ and $E_r \propto r^{-2}$.
%
%\begin{note}
%Insert figure of $T_{\mathrm{rad}}$ vs. $r$ and $E_r$ vs. $r$ for the optically thin irradiation of a uniform density medium example. Beside the temperature or radiation energy, draw offset straight line segments that are $\propto r^{-1/2}$ and $\propto r^{-2}$, respectively.
%\end{note}

\section{Irradiation of a static disk} \label{sec:static_disk_test}

A radiation test of general astrophysical interest is that of a star embedded in a circumstellar disk. The disk itself is optically thick and is surrounded by an optically thin envelope. The setup we use follows \citet{Pascucci+2004} and includes a sun-like star surrounded by a flared circumstellar disk similar to those considered by \citet{ChiangGoldreich97, ChiangGoldreich99} for T Tauri stars. The density structure features steep gradients in the inner disk, which can be challenging for radiative transfer codes, making this an excellent test.

We compare the temperature profile of the disk midplane in our simulation with one calculated using the radiative transfer module called \MAKEMAKE implemented in \PLUTO \citep{Mignone+2007} by \citet{Kuiper+2010b}. In the comparable setup ($\tau_{\mathrm{550nm}} \sim 100$), the difference between the simulated temperatures and Monte Carlo calculation was $\lesssim 16\%$. However, in the \citet{Pascucci+2004} benchmark paper, the temperature variation between different Monte Carlo codes including isotropic scattering is of the same order ($\lesssim 15\%$). 

The setup of the flared disk is as follows:
 
\begin{equation}
\rho(r,z) = \rho_0 \frac{r_d}{r} \exp\left(-\frac{\pi}{4}\left(\frac{z}{h(r)}\right)^2\right),
\end{equation}
where
\begin{equation}\label{eqn:scale_height}
h(r) = z_d \left(\frac{r}{r_d}\right)^{1.125},
\end{equation}
and

\begin{eqnarray}
r_d = \frac{r_{\mathrm{max}}}{2} = 500 \textrm{ AU} \\
z_d = \frac{r_{\mathrm{max}}}{8} = 125 \textrm{ AU}
\end{eqnarray}

We set up our computational domain to be a cube of side length 1000 AU, with the radiation source located at one corner. The source has radius $r = \Rsun$ and $T_{\mathrm{eft}} = 5800$ K. The domain represents one octant of a disk around a sunlike star. The minimum density is set to $\rho_{\mathrm{small}} = 10^{-23}$ g/cm$^3$ to avoid division-by-zero errors. We run simulations at two different fiducial densities $\rho_0$, representing the optically thin and optically thick cases. These runs are tabulated in Table \ref{tab:pascucci_sims}. We use the same fiducial densities as in \citet{Kuiper+2010b}, but our optical depths are computed for frequency-averaged (``gray'') radiation, considering only absorption and neglecting scattering, that is, $\tau = \tau_{\mathrm{abs}}$. In \citet{Kuiper+2010b} and \citet{Pascucci+2004}, $\tau = \tau_{\mathrm{abs}} + \tau_{\mathrm{sc}}$. When only absorption is considered, the optical depths in \MAKEMAKE align with our calculation \citetext{priv. comm.}.

Our simulation volume represents only one octant of the total protostellar disk, with the stellar source placed at one corner of the simulation volume. The masses tabulated in Table \ref{tab:pascucci_sims} are the total simulation masses multiplied by 8 so as to represent the total disk mass.

\begin{figure}
\includegraphics[width=88mm]{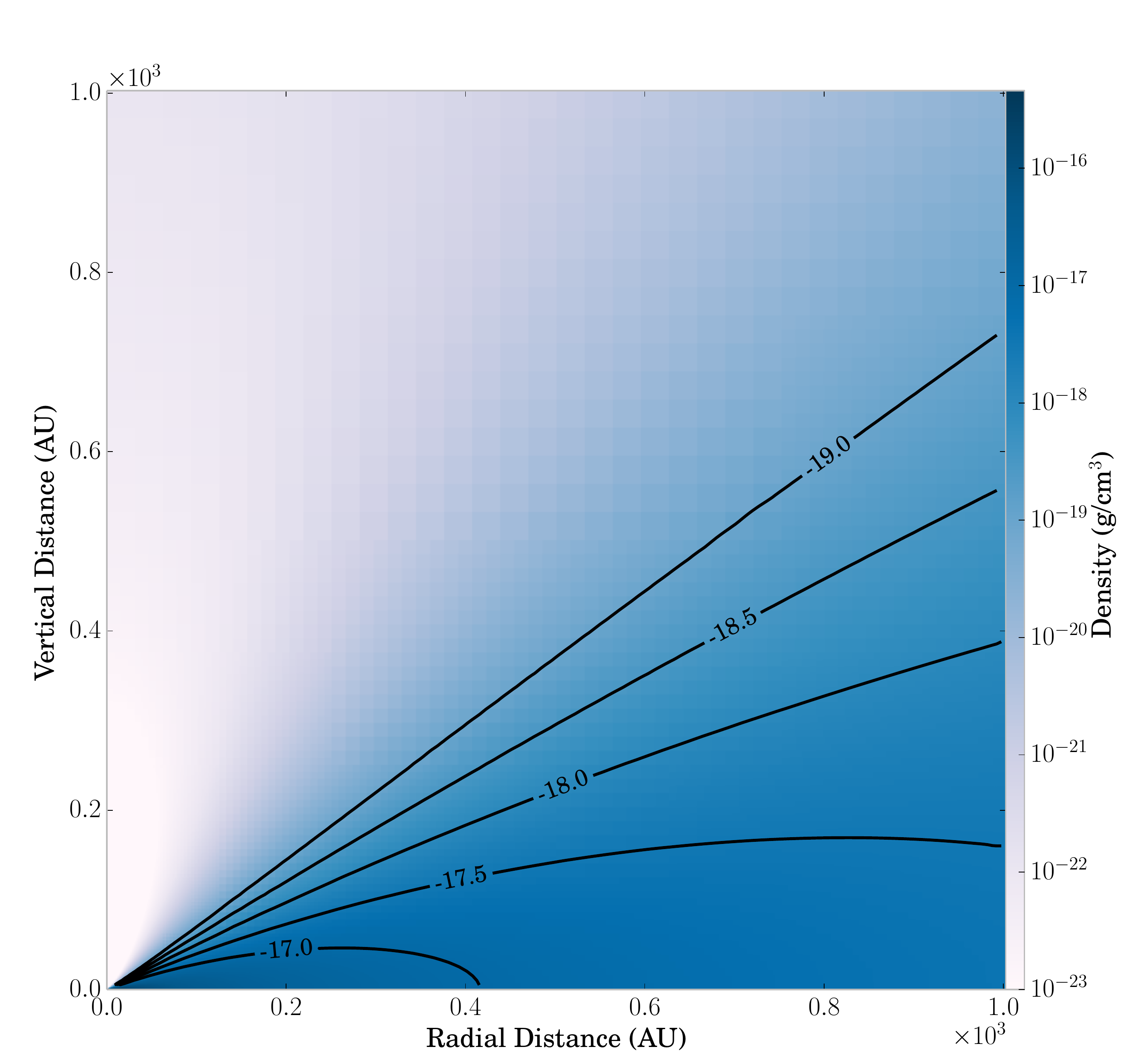}
\caption{Density profile of the \citet{Pascucci+2004} irradiated disk setup. Contours show lines of constant density. The radiation source is located at the bottom-left corner of the computational domain.}
\label{fig:pascucci_density}
\end{figure}

\begin{figure}
\includegraphics[width=88mm]{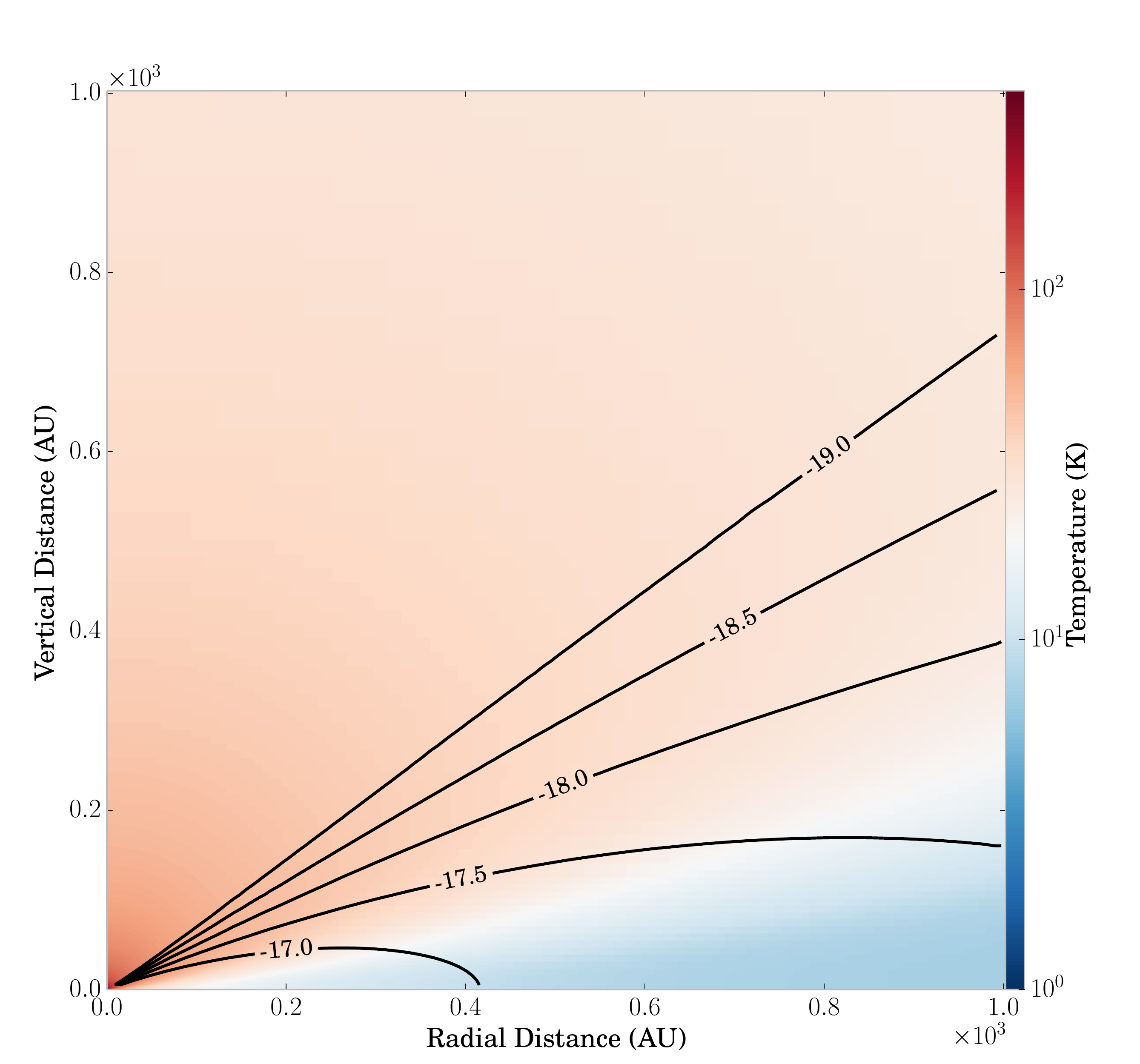}
\caption{Temperature profile of the \citet{Pascucci+2004} irradiated disk setup. Contours show lines of constant density. The gas is heated by a radiation source in the bottom-left corner of the computational domain. Note that the temperature has been scaled logarithmically.}
\label{fig:pascucci_temperature}
\end{figure}

\begin{figure}
\includegraphics[width=88mm]{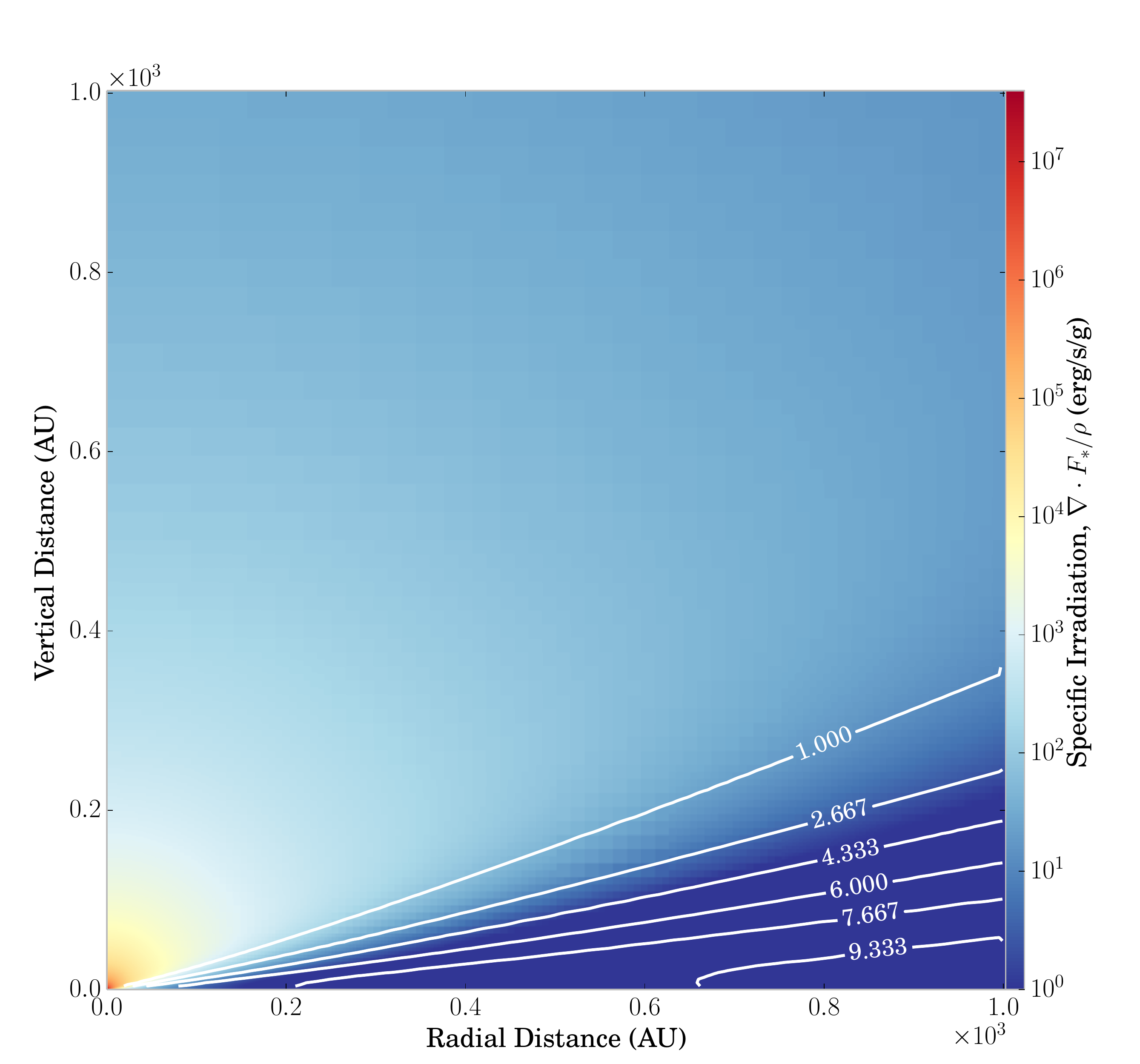}
\caption{Specific irradiation profile of the \citet{Pascucci+2004} irradiated disk setup, which shows the stellar radiation energy absorbed per gram of material. Contours show lines of constant optical depth, as computed by a raytrace through the material with frequency-averaged opacities.}
\label{fig:pascucci_irradiation}
\end{figure}

\begin{table}
  \centering
  \caption{Simulations of the irradiated disk setup.}
  \label{tab:pascucci_sims}
  \begin{tabular}{ccc}
    \hline
    $\tau_{\mathrm{gray,abs}}$ & $\rho_0$ [g cm$^{-3}$] & $M_{\mathrm{tot}}$ [$\Msun$] \\ 
    \hline
    0.01 & $8.321 \times 10^{-21}$ & $1.56 \times 10^{-5}$\\
    10.1 & $8.321 \times 10^{-18}$ & $1.56 \times 10^{-2}$\\
    \hline
  \end{tabular}
\end{table}

In the optically thin case, the temperature profile of the midplane can be compared to analytic estimates by \citet{Spitzer1978}. The gas/dust temperature far from the central star ($r \gg R_*$) is given by
\begin{equation}
T(r) = \left(\frac{R_{\mathrm{min}}}{2r}\right)^{\frac{2}{4+\beta}} T_{\mathrm{min}},
\end{equation}
where $\beta$ is the index of the dust absorption coefficient. For \citet{DraineLee1984} silicates, $\beta = 2.0508$ in the long wavelength regime \citep{KuiperKlessen2013}. $T_{\mathrm{min}}$ is the temperature at the inner edge of the disk $R_{\mathrm{min}} = 1$ AU.

In the optically thin regime, diffusion effects are negligible and the radiation field is dominated by the direct component. At radii greater than about 4 AU, our scheme reproduces the \citet{Spitzer1978} estimate to within 10\%, and for radii greater than 10 AU, better than 5\%.

Figure \ref{fig:pascucci_density} shows the density structure of the protostellar disk in a vertical slice in the optically thick case, with the radiation source positioned at the origin in the bottom-left corner. Contours mark lines of equal density and are labeled by the powers of ten of density.

The flared disk shields some of the stellar radiation. We run the simulation until the temperature of gas reaches an equilibrium state. Without hydrodynamics enabled, the gas can only absorb or radiate energy. We run the simulation for $10^{12}$ seconds and show the equilibrium temperature in figure \ref{fig:pascucci_temperature}.

To show the shielding properties of the disk and the energy being absorbed by the gaseous medium surrounding the star, we show the specific irradiation in figure \ref{fig:pascucci_irradiation}, that is, the stellar (direct) radiation energy being absorbed per gram of material. We therefore show lines of equal optical depth $\tau$, as calculated during the raytrace.

\begin{figure}
\includegraphics[width=88mm]{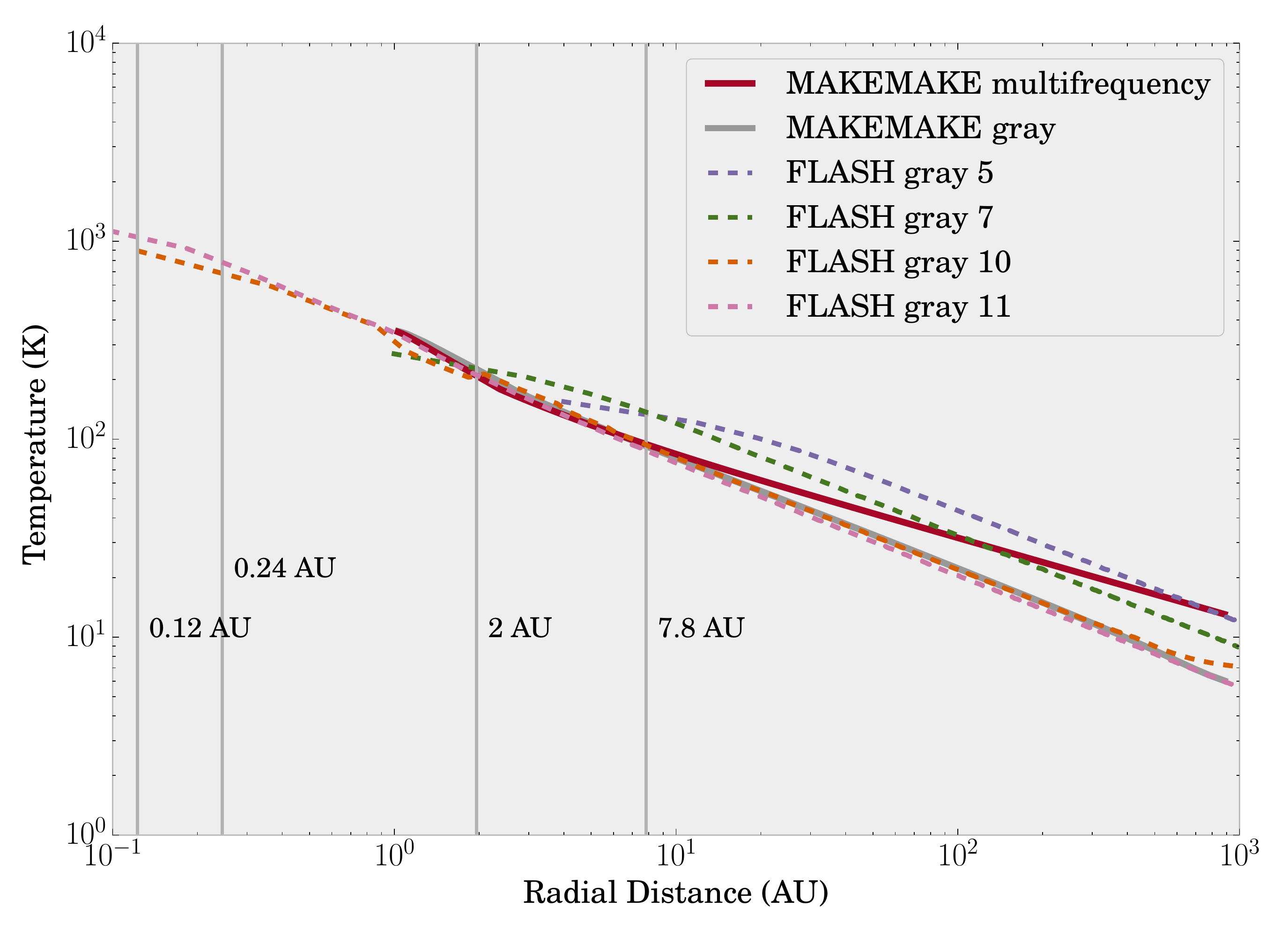}
\caption{Comparison of the midplane temperature profiles in simulations of the hybrid radiation feedback scheme in the \citet{Pascucci+2004} benchmark test in the high-density simulation ($\rho_0 = 8.321 \times 10^{-18}$): The implementation in \FLASH vs.\ \MAKEMAKE by \citet{Kuiper+2010b}. Solid lines belong to simulations using the \MAKEMAKE code, completed using a spherical polar geometry, while dashed lines indicate simulations done using the scheme described in this paper using the \FLASH code. The number associated with each \FLASH run indicates the maximum refinement level. Vertical gray lines in the figure indicate the grid resolution (in AU) for each of the \FLASH runs.} 
\label{fig:tau100_compare}
\end{figure}

We compare the temperatures through the midplane of the disk against the same calculation completed with the \MAKEMAKE code by \citet{Kuiper+2010b}. Their simulation was done in a spherical polar geometry, which, although not adaptively refined, naturally has more resolution in the polar angular component at small radii, and therefore the scale height of the disk is extremely well resolved.

Figure \ref{fig:tau100_compare} shows the radial temperature profile through the midplane of the disk for the case with fiducial density $\rho_0 = 8.321 \times 10^{18}$ g/cm$^3$. Three \FLASH runs are compared to two \MAKEMAKE simulations. The \FLASH simulations use the hybrid scheme as described in this paper, with both gray raytracer and gray FLD solver. These we compare against gray and multifrequency simulations by \citet{Kuiper+2010b}.

Because the \FLASH simulations were completed in a Cartesian AMR geometry, we indicate the maximum refinement level in the figure legend. We compare maximum levels 5, 7, 10, and 11. At 10 levels of refinment, the smallest grid size is $0.24$ AU; at 11 levels, it is $0.12$ AU. For this simulation to match the results obtained using \MAKEMAKE, it is necessary to resolve the scale height (Eq. \ref{eqn:scale_height}) of the disk at its inner edge ($\approx 0.1$ AU), which we effectively do with 11 levels of refinement, although convergence is already seen with 10 levels of refinment. When the scale height is resolved, the midplane temperature profile converges to the gray radiation result of \citet{Kuiper+2010b}. Lower resolution runs show a temperature excess.

\subsection{Radiation pressure on a static, flared disk} \label{sec:radiation_pressure_test}

\begin{figure}
\includegraphics[width=88mm]{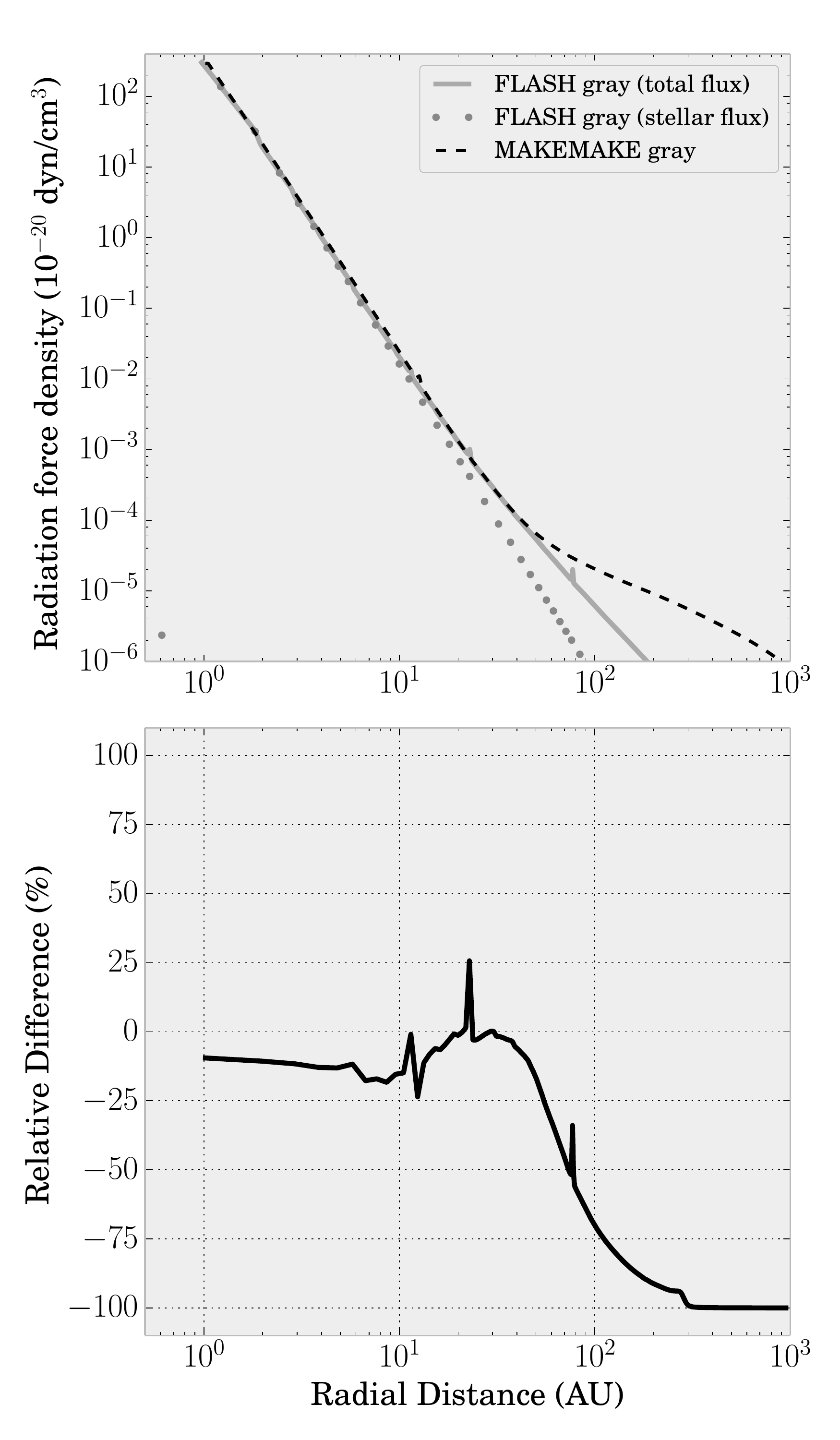}
\caption{Radiation force density through the midplane of the disk. The solid line indicates the radiation force density for the \FLASH calculation, resulting from the total flux (direct stellar and thermal). The dotted line indicates the radiation force density due only to the stellar component of the radiation field. We compare the \FLASH results to a 2T gray calculation in \MAKEMAKE (dashed line). The lower panel indicates the relative difference between the total flux radiation force density in \FLASH versus \MAKEMAKE.}
\label{fig:pascucci_frad}
\end{figure}

We compute the radiation force density for the stellar radiation component through the midplane of our disk as in equation \ref{eqn:radiation_force_density}. A similar test was done by \citet{Kuiper+2010b}, although their cut was not done through the midplane but along a polar angle of $\theta \approx 27^{\circ}$. \FLASH computes an isotropic radiation pressure via the Eddington approximation (equation \ref{eqn:eddington_approximation}), so we estimate the thermal radiation force density via
\begin{equation}
f_{\mathrm{rad,thermal}} = - \frac{d p_{\mathrm{rad}}}{dx}.
\end{equation}

We use the \MAKEMAKE code (revised to be a 2-T solver) to perform the same calculation through the midplane ($\theta \approx 0^{\circ}$) and compare to \FLASH. The results are shown in figure \ref{fig:pascucci_frad}. Most of the radiation pressure is due to the direct, stellar component, with the diffuse, reprocessed field adding only a tiny contribution that is more apparent far from the source. The outer boundary is treated as optically thin outflow in \MAKEMAKE. In \FLASH, we set the ``vacuum'' outer boundary condition, $dF/dx = -2F$.

In the inner 40 AU, the two codes match to within about 20\%, despite the radiation force density varying over 6 orders of magnitude. Because the radiation force is proportional to the absorbed radiation flux, the peak in the force density lies at the inner edge of the disk around 1 AU. Beyond 40 AU, the relative difference between the two calculations grows on account of boundary conditions, but the radiation force density is negligible beyond this point (6 orders of magnitude below the peak value). 

The \FLASH code and updated \MAKEMAKE code produce comparable measurements of the radiation pressure. The advantage and key innovation with \FLASH, however, is its AMR-capability that allows it to solve more general problems with high accuracy.

\section{Tests involving multiple sources} \label{sec:multiple_sources_tests}

Our radiative transfer scheme can accommodate multiple sources in a relatively straightforward way. We must calculate the direct radiation flux from stellar sources in order to compute the irradiation source term, $\grad \cdot \vec{F}_*$, at every cell in the computational domain. Therefore, we perform a loop over all sources and sum their flux contributions at each cell.

\subsection{Two proximal sources in a homogeneous medium} \label{sec:two_sources}

\begin{figure}
\includegraphics[width=88mm]{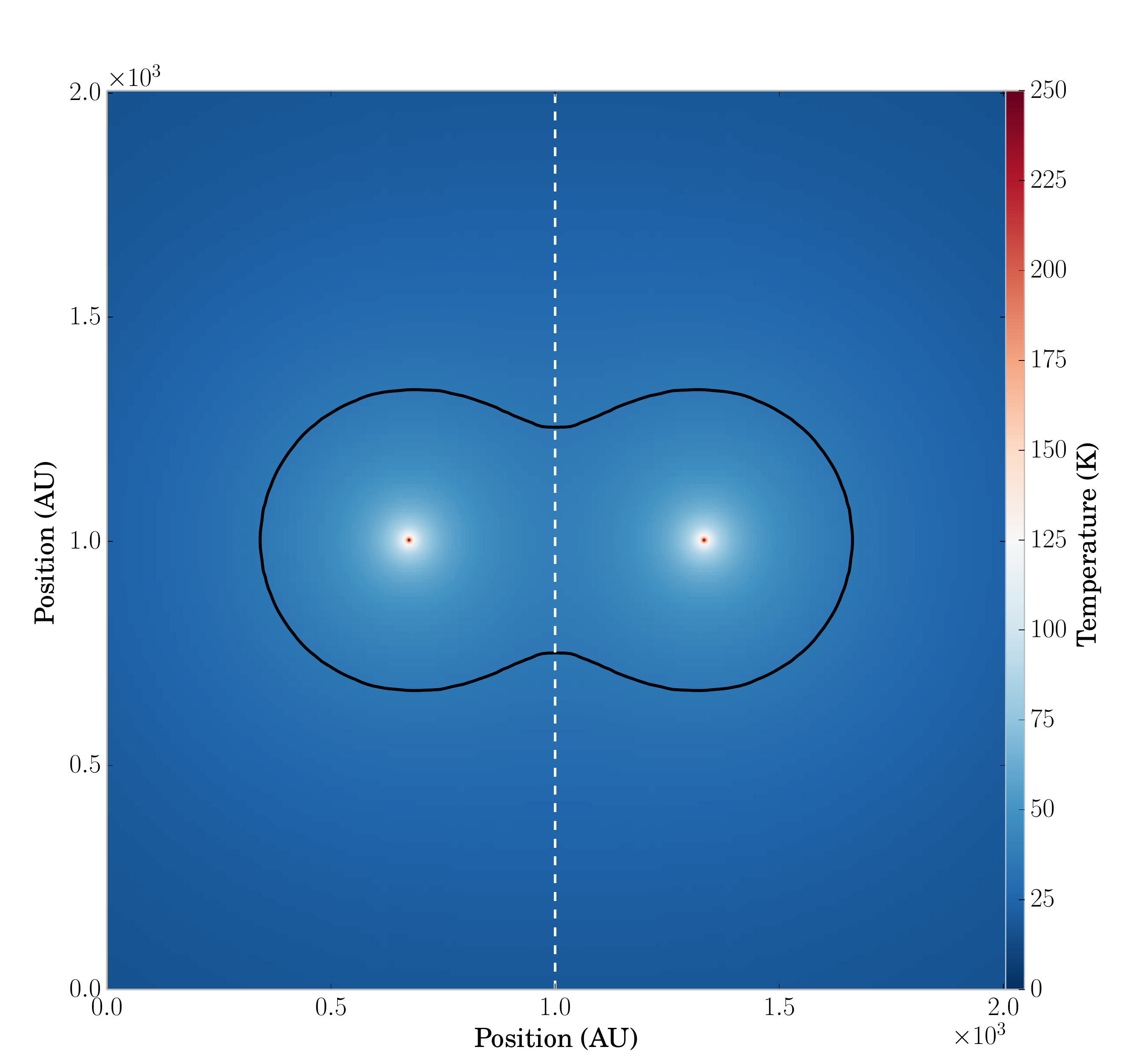}
\caption{Gas temperature after equilibrating with the radiation field driven by two stellar sources ($T = 5800$ K). Vertical dashed line indicates the symmetry axis along which the temperature in figure \ref{fig:two_source_midplane_temperature} is measured. The black contour marks a gas temperature of $T = 35$K.}
\label{fig:two_source_temperature}
\end{figure}

The first multi-source test involves two sources separated by about 658 AU. Each source has a stellar radiation field with $T_{\mathrm{eff}} = 5800$K. The medium is uniform and homogeneous, with a density of $\rho = 8.321 \times 10^{-18}$ g/cm$^3$. We perform this test to contrast our hybrid radiation transfer method with M1 moment methods \citep{Levermore1984,Gonzalez+2007,Vaytet+2010}. M1 moment methods are similar to FLD methods in that they are both based on taking angular moments of the radiative transfer equation. FLD takes only the zeroth-order moment, and the conservation equations are closed using a diffusion relation based on the gradient of the radiation energy. M1 methods take the first moment of the radiative transfer equation, and the closure of the conservation equations takes a form that preserves the bulk directionality of photon flows. Problems arise, however, when multiple sources are present. With two sources, photons flow in opposing directions, canceling the opposing components of their flow and introducing spurious flows in the direction perpendicular to the line between the two sources \citep{Rosdahl+2013}. Raytracers do not suffer from this artifact because the irradiation at each grid location is calculated along sightlines to each of the sources present in the simulation. 

\begin{figure}
\includegraphics[width=88mm]{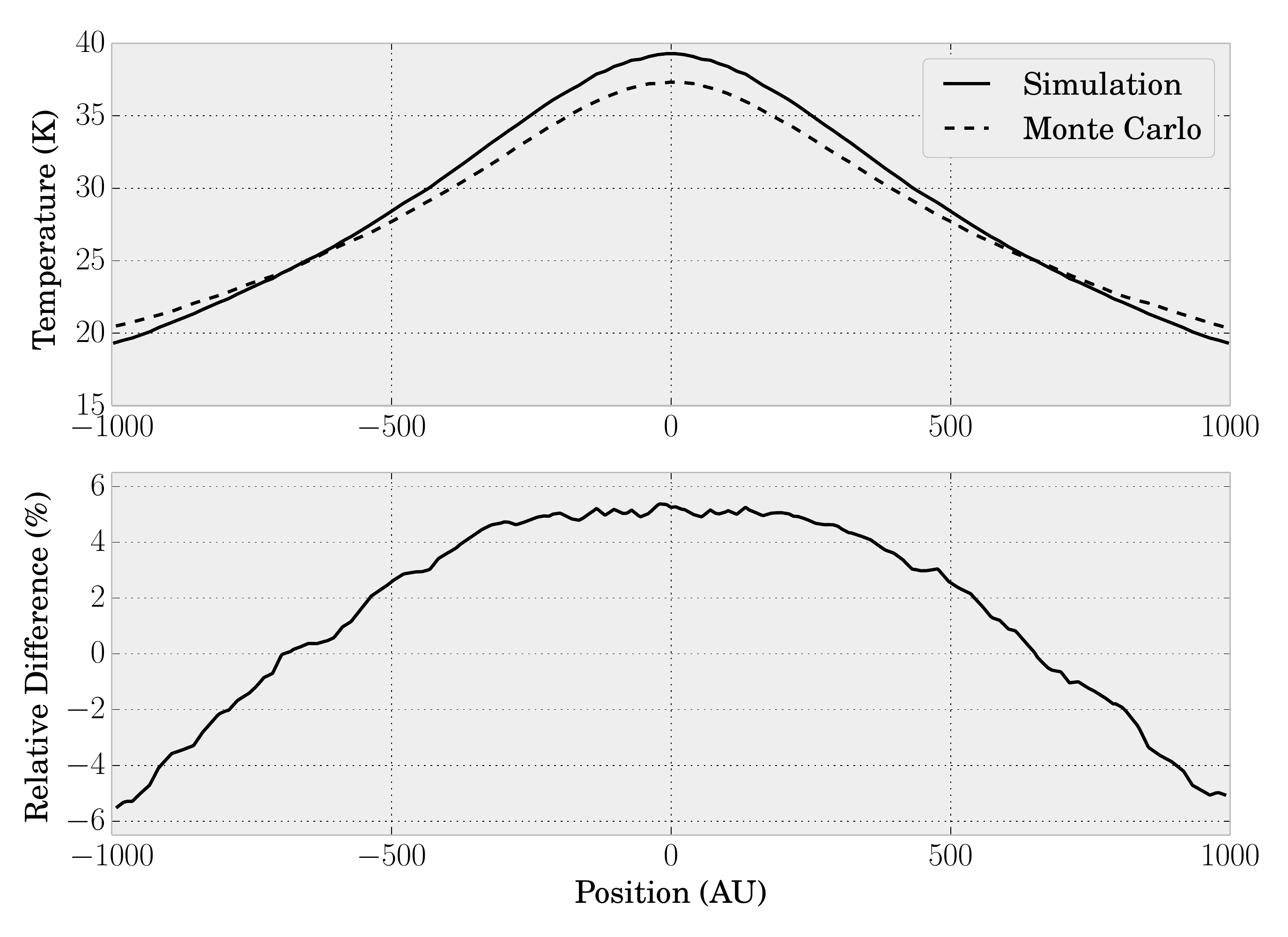}
\caption{Gas temperature along dashed line from figure \ref{fig:two_source_temperature}, the locus of points equidistant from both radiation sources. The upper panel shows the gas temperature from our simulation (solid) compared to the Monte Carlo result (dashed). The lower panel shows the relative difference in temperature as a percentage.}
\label{fig:two_source_midplane_temperature}
\end{figure}

Figure \ref{fig:two_source_temperature} shows the simulation setup as well as the temperature of the gas after it has been allowed to equilibrate with the radiation field. We measure the temperature of the gas along the vertical dashed line and plot the result in figure \ref{fig:two_source_midplane_temperature}.

Figure \ref{fig:two_source_midplane_temperature} shows the temperature along the dashed line from figure \ref{fig:two_source_temperature}, the locus of points equidistant from both radiation sources. Here we compare the simulated temperatures of our hybrid scheme in \FLASH against a Monte Carlo calculation performed with RADMC-3D\footnote{\url{http://www.ita.uni-heidelberg.de/~dullemond/software/radmc-3d/}} \citep{RADMC3D}. The Monte Carlo calculation was done with $10^{9}$ photons using the ``modified random walk'' method with scattering neglected, on an uneven grid (101x1x101 cells for the $x \times y \times z$ dimensions) using the same \citet{DraineLee1984} dust model as in the rest of this paper.

The \FLASH temperatures differ from the Monte Carlo results by less than 5\%, with \FLASH calculating slightly warmer temperatures directly between the sources and slightly cooler temperatures at the edges of the simulation volume. These small differences are most likely due to multifrequency effects.

Monte Carlo calculations are considered the benchmark for accuracy, but are extremely computationally intensive and cannot, in general, be used in dynamic calculations. Simulations of star formation cannot be addressed with full Monte Carlo methods using present-day machines and reasonable time constraints. Moreover, they are highly nontrivial to parallelize. In these cases, approximate schemes such as flux-limited diffusion must be used.

\subsection{Two sources irradiating a dense core of material} \label{sec:shadow_test}

We next create a test setup inspired by \citet{Rijkhorst+2006}, where a central concentration was irradiated by two sources. Since FLD methods are based on local gradients in the radiation energy, they cannot properly treat shadowing. In this test we place a dense clump of material at the center of our simulation volume and irradiate it from two angles, creating a shadowed region. We compute the irradiation and equilibrium temperatures.

Our setup contains a mix of optically thin and optically thick regions, creating strong gradients in the radiation energy density. We compare the equilibrium temperatures computed by the hybrid scheme to the temperature computed without diffusion.

The two radiation sources have a temperature $T_{\mathrm{eff}} = 8000$ K and are situated in a low-density ($\rho = 10^{-20}$ g/cm$^{-3}$) medium to the side and below a central density concentration ($\rho = 10^{-17}$ g/cm$^{-3}$) with radius $R_c = 4 \times 10^{15}$ cm $\approx 267$ AU. The side length of the simulation box is about 2000 AU. Hydrodynamics are disabled.

\begin{figure}
\includegraphics[width=88mm]{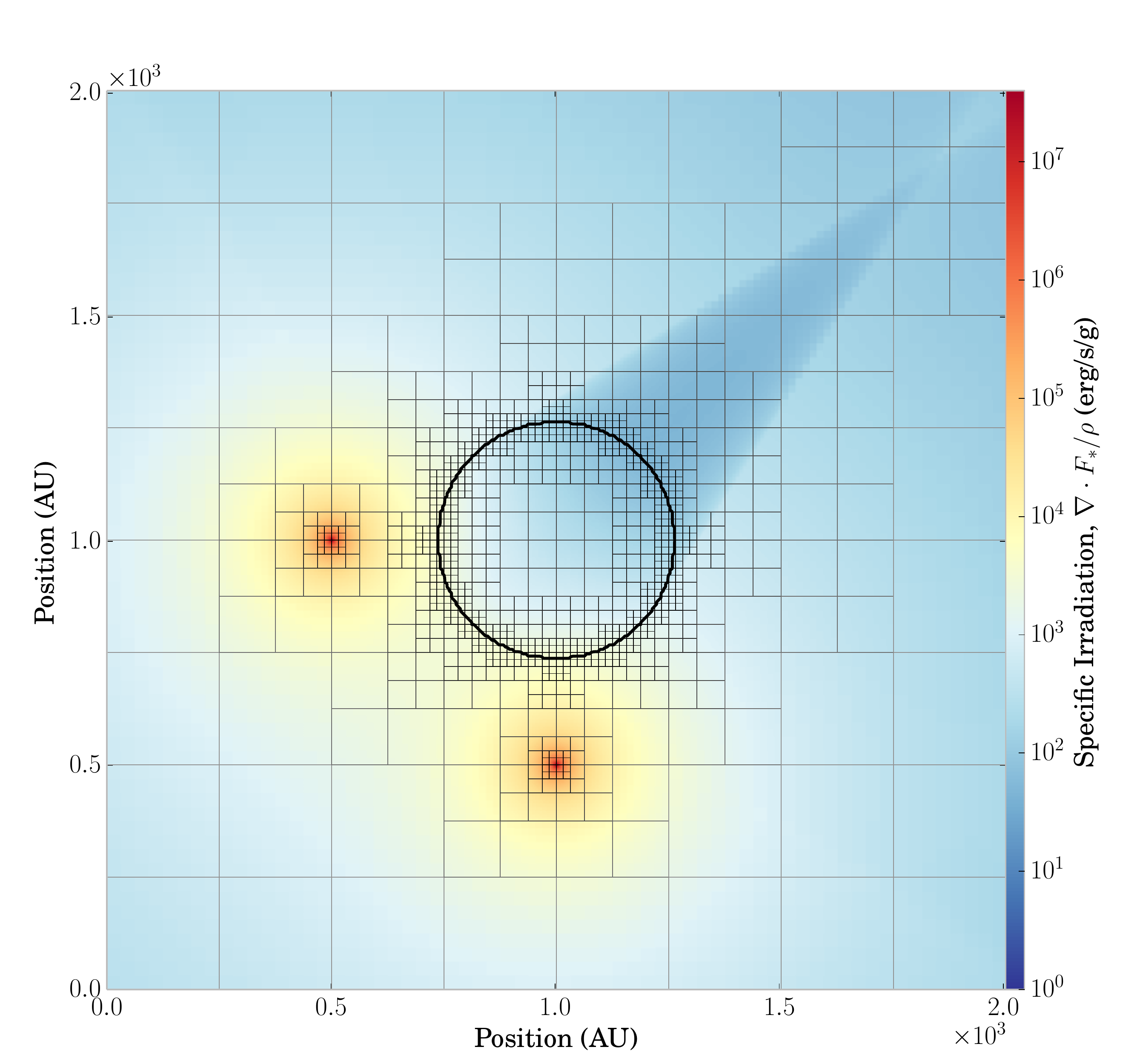}
\caption{Specific irradiation by two sources in an otherwise homogeneous medium at $\rho = 10^{-20}$ g/cm$^3$, but for a central density concentration indicated by the black circle, where the density is $\rho = 10^{-17}$ g/cm$^3$. The two sources are stellar sources with effective temperatures of 8000K. The \FLASH block AMR structure is also shown, with each block containing $8^3$ cells.}
\label{fig:shadow_irradiation}
\end{figure}

Figure \ref{fig:shadow_irradiation} shows the simulation setup and the specific irradiation of the gas in a slice through the midplane of the simulation volume. The overlaid grid shows the \FLASH AMR block structure, with each block containing $8^3$ cells and refining on gradients in density, radiation energy, and matter temperature using the \FLASH error estimator described in section \ref{sec:FLASH}. The simulation was run with a maximum 8 levels of refinement, achieving a maximum resolution of 1.96 AU. The black circle in the centre marks the central density concentration. The central density concentration casts a shadow on the side opposite of the central concentration.

\begin{figure}
\includegraphics[width=88mm]{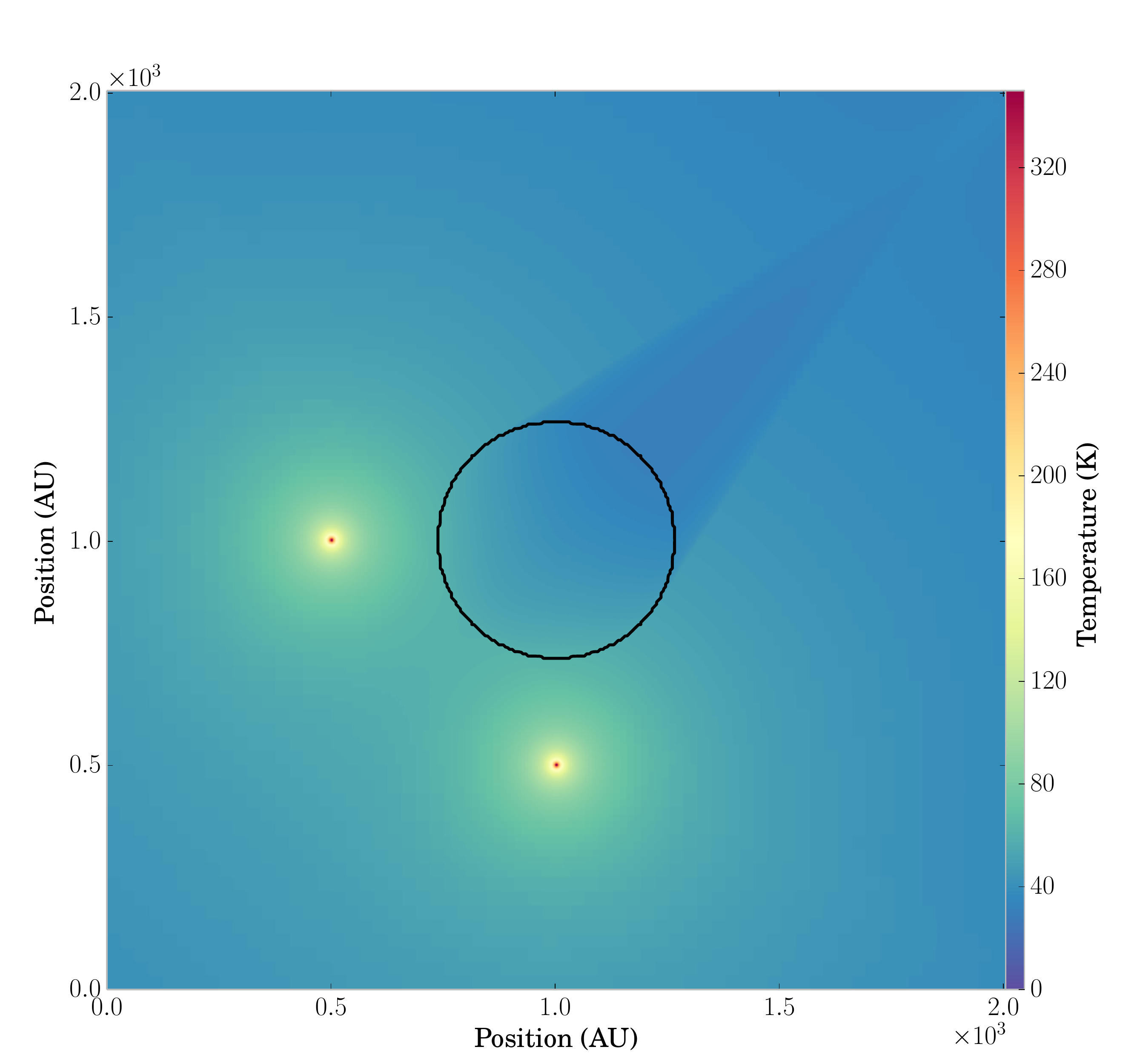}
\caption{Same as in figure \ref{fig:shadow_irradiation}, but instead showing the gas temperature.}
\label{fig:shadow_temperature}
\end{figure}

Figure \ref{fig:shadow_temperature} shows the gas temperature after the simulation has been given time to reach equilibrium. Here we see warming of a region just interior to the central overdensity where strong density gradients exist.

\begin{figure}
\includegraphics[width=88mm]{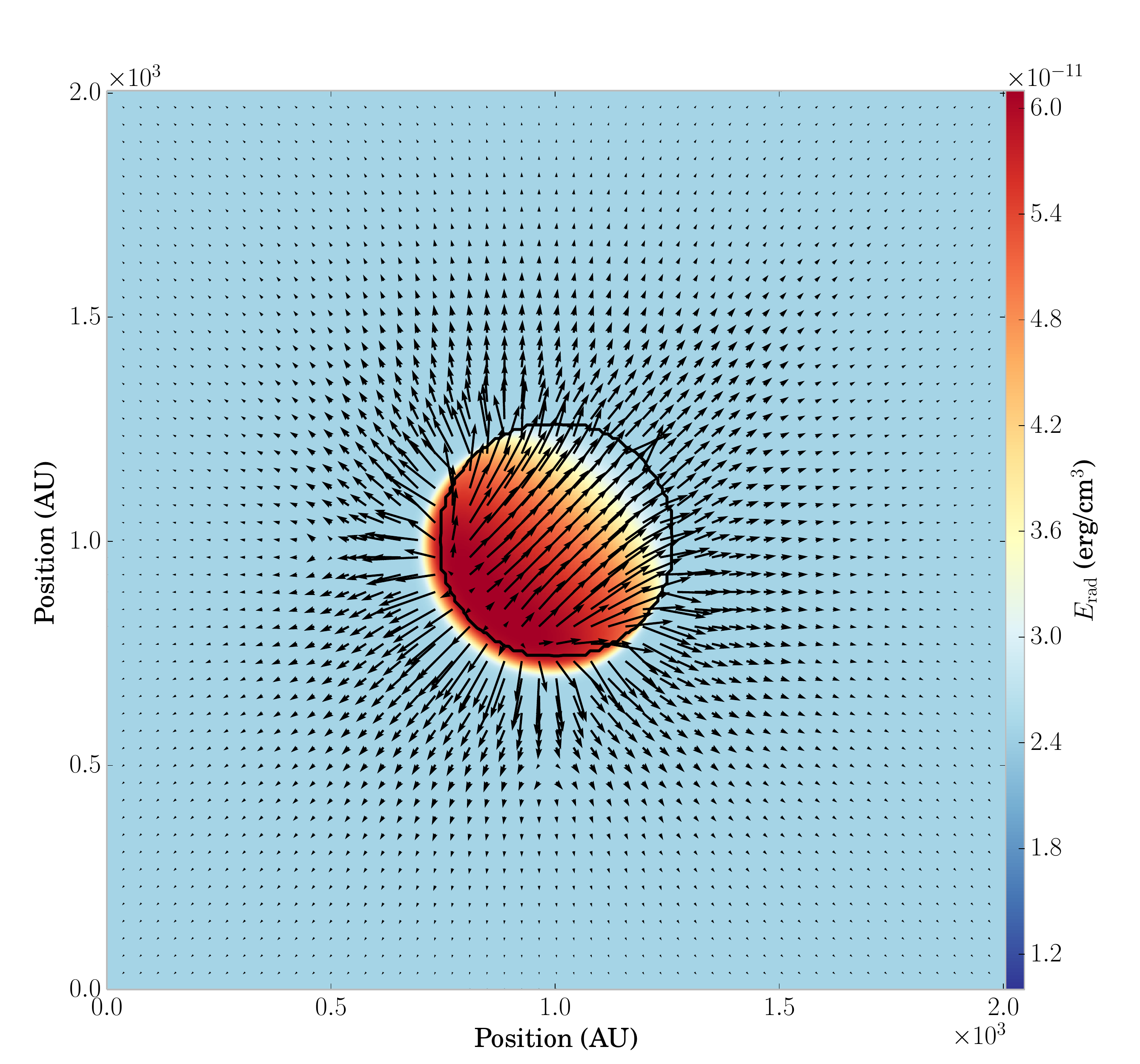}
\caption{Radiation energy density with the thermal radiation flux $\vec{F}_r = -D \grad E_r$ overplotted as a vector field. The gradients in the radiation energy density are steepest near the spherical overdensity in the centre, on the side facing the two sources.}
\label{fig:shadow_erad_gradients}
\end{figure}

To visualize the radiation flux, we plot a vector field of the radiation flux, given by equation \ref{eqn:fld_approximation}. This is shown in figure \ref{fig:shadow_erad_gradients} overplotted on the total radiation energy density. The vector field represents how the radiation energy is being diffused by the FLD solver. The two radiation sources irradiate the central overdensity, heating it. It then drives radiation back into the surrounding medium via diffusion. There is also a radiation energy gradient across the central overdensity. The black circle marks the central overdensity, as in the previous two figures. In figure \ref{fig:shadow_erad_gradients} we see a halo of radiation energy just outside the circle on the side nearest to the two radiation sources also resulting from the diffusion of radiation. 

\begin{figure}
\includegraphics[width=88mm]{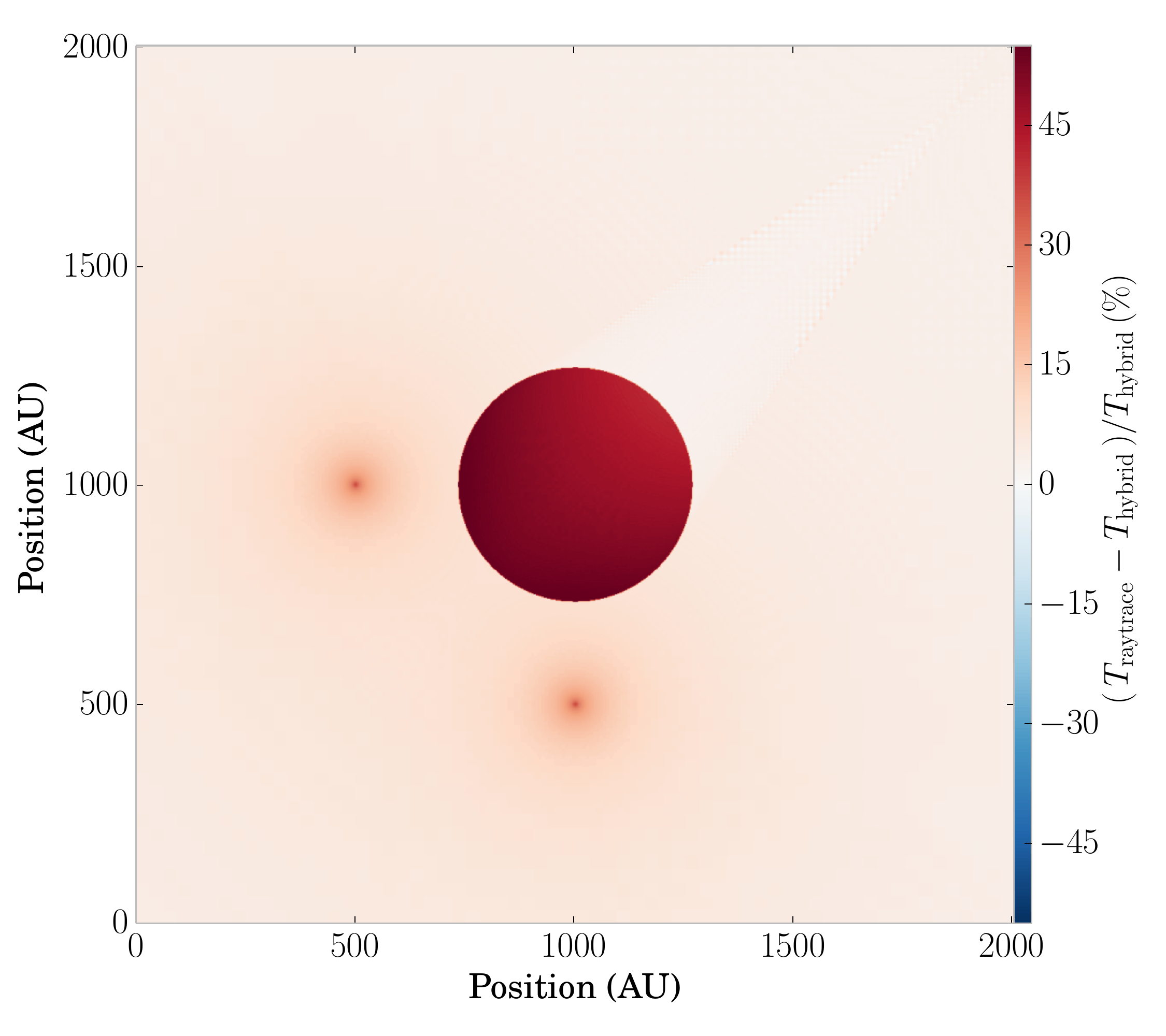}
\caption{The temperature difference (in percent) in the raytrace case compared to the hybrid radiation transfer case.}
\label{fig:compare_raytrace_hybrid}
\end{figure}

We now compare this to the zero-diffusion case by performing the same calculation with the diffusion term set to zero. The gas equilibrates with the radiation field imposed by the sources, but this energy is not allowed to diffuse. Figure \ref{fig:compare_raytrace_hybrid} shows the difference in the resultant equilibrium temperature as a percentage of the relative difference. The result is that the central overdensity is approximately 50\% hotter. It cannot cool by diffusing its radiation energy into the surrounding medium.

Because the re-emission terms are often left out of raytracing-methods for the sake of simplicity and computational cost, they are unsuitable and inaccurate in simulations containing optically thick material, although re-emission is often treated using a cooling function to extract the energy. In optically thick envelopes surrounding young stellar objects, radiation is reprocessed as the envelope of gas absorbs and re-emits the radiation in the infrared. 

\section{Discussion and Summary} \label{sec:conclusion}

We have described the implementation of a hybrid radiation transfer scheme in a 3D Cartesian AMR framework. To the best of our knowledge, this is the first such implementation. Our hybrid scheme splits the radiation field into a direct (stellar) component and a diffuse (thermal) component. A specialized raytracer is used to solve for the direct flux, the absorption of which by dusty, molecular gas constitutes an energy source term that heats the dust. The dust emits radiation thermally, and the thermal radiation field is evolved via flux-limited diffusion. The gas is assumed in thermal equilibrium with the dust, whose temperature is evolved according to both direct and indirect radiation fields.

Radiation feedback from massive stars takes manifold forms: heating of dust grains, which are coupled to the gas; radiation pressure; ionizing radiation, leading to the formation of \HII regions; jets and stellar winds; and finally supernovae. We have implemented the first two, with future work intent on implementing ionizing radiation. Various authors have drawn attention to the potential for ionizing radiation to disrupt molecular clouds \citep{Matzner2002,Krumholz2006,Walch+2012}, at least in low-mass clouds, while in larger clouds, star formation may be fed through accretion along filaments, with ionizing radiation feeding into the low-density voids between filaments \cite{Dale+2012}. The effect of ionizing radiation, through heating of the gas and suppression of fragmentation, also leads to higher Jeans masses and more massive stars \citep{Peters+2010c}.

As a mechanism for disrupting giant molecular clouds, \citet{Murray+2010} argue that radiation pressure from starlight interacting with dust grains plays a dominant role, at least in the case of the most massive clusters inside GMCs.

In implementing our radiative transfer scheme, the addition of the raytracer helps overcome some of the main limitation of a purely FLD approach. The primary limitation of the FLD approach is the assumption that the flux always travels in the direction down the radiation energy gradient. This assumption is reasonable in purely optically thick regions, but breaks down in regions where radiation sources ought to be casting shadows. 

Furthermore, some FLD implementations assume that the radiation field is everywhere in thermal equilibrium with the gas. This is not always true, especially radiatively critical shocks. We have relaxed this assumption and allowed the radiation temperature and the gas temperature to differ from each other.

The primary advance over the hybrid implementation of \citet{Kuiper+2010b} is the generalization to 3D Cartesian AMR from the specialized geometry of a spherical polar grid. Spherical coordinates are ideal for studying the radiation field from a single, central source. By their nature, spherical grids have higher effective resolution closer to the center, precisely where it is most needed simulations of accretion disks around massive stars. In our test problem of a static, irradiated disk \citep{Pascucci+2004}, we required 11 levels of refinement before we could resolve the scale height of the disk near the inner boundary.

However, when generalizing to multiple point sources of radiation, Cartesian geometry becomes a natural choice. This enables us to approach a much larger array of problems. \FLASH is a highly scalable MHD code with driven turbulence, sink particles and protostellar evolution. The addition of our hybrid radiation transfer unit greatly complements \FLASH's abilities.

One of the most general problems this enables us to study is the effect of radiative heating and momentum feedback around massive, accreting protostars. These objects reside deep within dense envelopes and accrete material through a disk. Whether disk accretion is the only mechanism, or whether raditiave Rayleigh-Taylor instabilities are another viable mechanism has not yet been fully settled. High resolution simulations, resolving in particular the regions of intermediate optical depth using adaptive mesh refinment, and including the effects of turbulence and MHD ought to be able to resolve the remaining controversies.

We have also connected the hybrid radiation scheme to sink particles, which we will use to model the formation of protostars in larger-scale simulations of molecular cloud clumps. The protostellar properties such as stellar radius and luminosity are evolved using a subgrid model \citep{Klassen+2012a} that directly feeds into the radiation transfer subroutines. The sum of \FLASH's capabilities now enable the study of clustered star formation in turbulent, filamentary, and magnetized environments.

The code also presents opportunities for further development. The raytracer, as implemented by \citet{Rijkhorst+2006} and \citet{Peters+2010a}, contained ionization feedback, which we will soon implement in our hybrid radiation framework. Multifrequency effects will also be included in future versions of the code.

\section*{Acknowledgments} \label{sec:acknolwedgments}

The development of this hybrid radiation transfer code benefited from helpful discussions with Romain Teyssier, Christoph Federrath, Klaus Weide, and Petros Tzeferacos, and from two visits by M.K.~to the \FLASH Center at the University of Chicago. M.K.~acknowledges financial support from the National Sciences and Engineering Research Council (NSERC) of Canada through a Postgraduate Scholarship. R.K.~acknowledges funding from the Max Planck Research Group \textit{Star Formation throughout the Milky Way Galaxy} at the Max Planck Institute for Astronomy. R.E.P.~is supported by an NSERC Discovery Grant. T.P. acknowledges financial support through a Forschungskredit of the University of Z\"{u}rich, grant no. FK-13-112. The \FLASH code was in part developed by the DOE-supported Alliances Center for Astrophysical Thermonuclear Flashes (ASCI) at the University of Chicago. This work was made possible by the facilities of the Shared Hierarchical Academic Research Computing Network (SHARCNET: www.sharcnet.ca) and Compute/Calcul Canada.

We are grateful to KITP, Santa Barbara, for supporting M.K. as an affiliate (for 3 weeks), and R.E.P. as an invited participant (for 2.5 months) in the program ``Gravity's Loyal Opposition: The Physics of Star Formation Feedback,'' where some of this research was supported by the National Science Foundation under Grant No. NSF PHY11-25915. 

Much of the analysis and data visualization was performed using the {\tt yt} toolkit\footnote{\url{http://yt-project.org}} by \citet{ytpaper}. 

\bibliography{hybrid_radiation_scheme}

\begin{thebibliography}{79}
\expandafter\ifx\csname natexlab\endcsname\relax\def\natexlab#1{#1}\fi

\bibitem[{{Agertz} {et~al.}(2013){Agertz}, {Kravtsov}, {Leitner}, \&
  {Gnedin}}]{Agertz+2013}
{Agertz}, O., {Kravtsov}, A.~V., {Leitner}, S.~N., \& {Gnedin}, N.~Y. 2013,
  \apj, 770, 25

\bibitem[{{Aikawa} \& {Herbst}(1999)}]{AikawaHerbst1999}
{Aikawa}, Y., \& {Herbst}, E. 1999, \aap, 351, 233

\bibitem[{{Banerjee} {et~al.}(2006){Banerjee}, {Pudritz}, \&
  {Anderson}}]{Banerjee+2006}
{Banerjee}, R., {Pudritz}, R.~E., \& {Anderson}, D.~W. 2006, \mnras, 373, 1091

\bibitem[{{Beuther} {et~al.}(2007){Beuther}, {Churchwell}, {McKee}, \&
  {Tan}}]{Beuther+2007}
{Beuther}, H., {Churchwell}, E.~B., {McKee}, C.~F., \& {Tan}, J.~C. 2007,
  Protostars and Planets V, 165

\bibitem[{{Bodenheimer} {et~al.}(1990){Bodenheimer}, {Yorke}, {Rozyczka}, \&
  {Tohline}}]{Bodenheimer+1990}
{Bodenheimer}, P., {Yorke}, H.~W., {Rozyczka}, M., \& {Tohline}, J.~E. 1990,
  \apj, 355, 651

\bibitem[{{Chiang} \& {Goldreich}(1997)}]{ChiangGoldreich97}
{Chiang}, E.~I., \& {Goldreich}, P. 1997, \apj, 490, 368

\bibitem[{{Chiang} \& {Goldreich}(1999)}]{ChiangGoldreich99}
---. 1999, \apj, 519, 279

\bibitem[{{Colella} \& {Woodward}(1984)}]{ColellaWoodward1984}
{Colella}, P., \& {Woodward}, P.~R. 1984, Journal of Computational Physics, 54,
  174

\bibitem[{{Commer{\c c}on} {et~al.}(2011){Commer{\c c}on}, {Teyssier}, {Audit},
  {Hennebelle}, \& {Chabrier}}]{Commercon+2011}
{Commer{\c c}on}, B., {Teyssier}, R., {Audit}, E., {Hennebelle}, P., \&
  {Chabrier}, G. 2011, \aap, 529, A35

\bibitem[{{Crutcher}(2012)}]{Crutcher2012}
{Crutcher}, R.~M. 2012, \araa, 50, 29

\bibitem[{{Dale} {et~al.}(2012){Dale}, {Ercolano}, \& {Bonnell}}]{Dale+2012}
{Dale}, J.~E., {Ercolano}, B., \& {Bonnell}, I.~A. 2012, \mnras, 424, 377

\bibitem[{{Draine} \& {Lee}(1984)}]{DraineLee1984}
{Draine}, B.~T., \& {Lee}, H.~M. 1984, \apj, 285, 89

\bibitem[{Dubey {et~al.}(2009)Dubey, Antypas, Ganapathy, Reid, Riley, Sheeler,
  Siegel, \& Weide}]{Dubey+2009}
Dubey, A., Antypas, K., Ganapathy, M.~K., {et~al.} 2009, Parallel Computing,
  35, 512

\bibitem[{{Dullemond}(2012)}]{RADMC3D}
{Dullemond}, C.~P. 2012, {RADMC-3D: A multi-purpose radiative transfer tool},
  astrophysics Source Code Library

\bibitem[{{Edgar} \& {Clarke}(2003)}]{EdgarClarke2003}
{Edgar}, R., \& {Clarke}, C. 2003, \mnras, 338, 962

\bibitem[{{Ensman}(1994)}]{Ensman1994}
{Ensman}, L. 1994, \apj, 424, 275

\bibitem[{{Evans} {et~al.}(2009){Evans}, {Dunham}, {J{\o}rgensen}, {Enoch},
  {Mer{\'{\i}}n}, {van Dishoeck}, {Alcal{\'a}}, {Myers}, {Stapelfeldt},
  {Huard}, {Allen}, {Harvey}, {van Kempen}, {Blake}, {Koerner}, {Mundy},
  {Padgett}, \& {Sargent}}]{Evans+2009}
{Evans}, II, N.~J., {Dunham}, M.~M., {J{\o}rgensen}, J.~K., {et~al.} 2009,
  \apjs, 181, 321

\bibitem[{Falgout \& Yang(2002)}]{HYPRE}
Falgout, R., \& Yang, U. 2002, in Lecture Notes in Computer Science, Vol. 2331,
  Computational Science --€" ICCS 2002, ed. P.~Sloot, A.~Hoekstra, C.~Tan, \&
  J.~Dongarra (Springer Berlin Heidelberg), 632--641

\bibitem[{{Fatenejad} {et~al.}(2012){Fatenejad}, {Bachan}, {Couch}, {Daley},
  {Dubey}, {Flocke}, {Graziani}, {Lamb}, {Lee}, {Scopatz}, {Tzeferacos}, \&
  {Weide}}]{Fatenejad+2012}
{Fatenejad}, M., {Bachan}, J., {Couch}, S., {et~al.} 2012, in APS Meeting
  Abstracts, 8081P

\bibitem[{{Federrath} {et~al.}(2010){Federrath}, {Banerjee}, {Clark}, \&
  {Klessen}}]{Federrath+2010}
{Federrath}, C., {Banerjee}, R., {Clark}, P.~C., \& {Klessen}, R.~S. 2010,
  \apj, 713, 269

\bibitem[{{Flock} {et~al.}(2013){Flock}, {Fromang}, {Gonz{\'a}lez}, \&
  {Commer{\c c}on}}]{Flock+2013}
{Flock}, M., {Fromang}, S., {Gonz{\'a}lez}, M., \& {Commer{\c c}on}, B. 2013,
  \aap, 560, A43

\bibitem[{{Fogel} {et~al.}(2011){Fogel}, {Bethell}, {Bergin}, {Calvet}, \&
  {Semenov}}]{Fogel+2011}
{Fogel}, J.~K.~J., {Bethell}, T.~J., {Bergin}, E.~A., {Calvet}, N., \&
  {Semenov}, D. 2011, \apj, 726, 29

\bibitem[{{Fryxell} {et~al.}(2000){Fryxell}, {Olson}, {Ricker}, {Timmes},
  {Zingale}, {Lamb}, {MacNeice}, {Rosner}, {Truran}, \& {Tufo}}]{Fryxell+2000}
{Fryxell}, B., {Olson}, K., {Ricker}, P., {et~al.} 2000, \apjs, 131, 273

\bibitem[{{Gonz{\'a}lez} {et~al.}(2007){Gonz{\'a}lez}, {Audit}, \&
  {Huynh}}]{Gonzalez+2007}
{Gonz{\'a}lez}, M., {Audit}, E., \& {Huynh}, P. 2007, \aap, 464, 429

\bibitem[{{Hayes} \& {Norman}(2003)}]{HayesNorman2003}
{Hayes}, J.~C., \& {Norman}, M.~L. 2003, \apjs, 147, 197

\bibitem[{{Klassen} {et~al.}(2012){Klassen}, {Pudritz}, \&
  {Peters}}]{Klassen+2012a}
{Klassen}, M., {Pudritz}, R.~E., \& {Peters}, T. 2012, \mnras, 421, 2861

\bibitem[{{Kolb} {et~al.}(2013){Kolb}, {Stute}, {Kley}, \&
  {Mignone}}]{Kolb+2013}
{Kolb}, S.~M., {Stute}, M., {Kley}, W., \& {Mignone}, A. 2013, \aap, 559, A80

\bibitem[{{Krumholz} {et~al.}(2007{\natexlab{a}}){Krumholz}, {Klein}, \&
  {McKee}}]{Krumholz+2007a}
{Krumholz}, M.~R., {Klein}, R.~I., \& {McKee}, C.~F. 2007{\natexlab{a}}, \apj,
  656, 959

\bibitem[{{Krumholz} {et~al.}(2007{\natexlab{b}}){Krumholz}, {Klein}, {McKee},
  \& {Bolstad}}]{Krumholz+2007b}
{Krumholz}, M.~R., {Klein}, R.~I., {McKee}, C.~F., \& {Bolstad}, J.
  2007{\natexlab{b}}, \apj, 667, 626

\bibitem[{{Krumholz} {et~al.}(2009){Krumholz}, {Klein}, {McKee}, {Offner}, \&
  {Cunningham}}]{Krumholz2009}
{Krumholz}, M.~R., {Klein}, R.~I., {McKee}, C.~F., {Offner}, S.~S.~R., \&
  {Cunningham}, A.~J. 2009, Science, 323, 754

\bibitem[{{Krumholz} {et~al.}(2006){Krumholz}, {Matzner}, \&
  {McKee}}]{Krumholz2006}
{Krumholz}, M.~R., {Matzner}, C.~D., \& {McKee}, C.~F. 2006, \apj, 653, 361

\bibitem[{{Krumholz} {et~al.}(2005){Krumholz}, {McKee}, \&
  {Klein}}]{Krumholz+2005b}
{Krumholz}, M.~R., {McKee}, C.~F., \& {Klein}, R.~I. 2005, \apjl, 618, L33

\bibitem[{{Krumholz} \& {Tan}(2007)}]{KrumholzTan2007}
{Krumholz}, M.~R., \& {Tan}, J.~C. 2007, \apj, 654, 304

\bibitem[{{Kuiper} {et~al.}(2010{\natexlab{a}}){Kuiper}, {Klahr}, {Beuther}, \&
  {Henning}}]{Kuiper+2010a}
{Kuiper}, R., {Klahr}, H., {Beuther}, H., \& {Henning}, T. 2010{\natexlab{a}},
  \apj, 722, 1556

\bibitem[{{Kuiper} {et~al.}(2011){Kuiper}, {Klahr}, {Beuther}, \&
  {Henning}}]{Kuiper+2011}
---. 2011, \apj, 732, 20

\bibitem[{{Kuiper} {et~al.}(2012){Kuiper}, {Klahr}, {Beuther}, \&
  {Henning}}]{Kuiper+2012}
---. 2012, \aap, 537, A122

\bibitem[{{Kuiper} {et~al.}(2010{\natexlab{b}}){Kuiper}, {Klahr}, {Dullemond},
  {Kley}, \& {Henning}}]{Kuiper+2010b}
{Kuiper}, R., {Klahr}, H., {Dullemond}, C., {Kley}, W., \& {Henning}, T.
  2010{\natexlab{b}}, \aap, 511, A81

\bibitem[{{Kuiper} \& {Klessen}(2013)}]{KuiperKlessen2013}
{Kuiper}, R., \& {Klessen}, R.~S. 2013, \aap, 555, A7

\bibitem[{{Kuiper} {et~al.}(2014){Kuiper}, {Yorke}, \&
  {Turner}}]{KuiperYorkeTurner2014}
{Kuiper}, R., {Yorke}, H., \& {Turner}, N.~J. 2014, submitted to ApJ

\bibitem[{{Kuiper} \& {Yorke}(2013{\natexlab{a}})}]{KuiperYorke2013a}
{Kuiper}, R., \& {Yorke}, H.~W. 2013{\natexlab{a}}, \apj, 763, 104

\bibitem[{{Kuiper} \& {Yorke}(2013{\natexlab{b}})}]{KuiperYorke2013b}
---. 2013{\natexlab{b}}, \apj, 772, 61

\bibitem[{{Kumar} {et~al.}(2011){Kumar}, {Bachan}, {Couch}, {Daley}, {Dubey},
  {Fatenejad}, {Flocke}, {Graziani}, {Lamb}, {Lee}, \& {Weide}}]{Kumar+2011}
{Kumar}, S., {Bachan}, J., {Couch}, S., {et~al.} 2011, in APS Meeting
  Abstracts, 9131P

\bibitem[{{Lamb} {et~al.}(2010){Lamb}, {Couch}, {Dubey}, {Gopal}, {Graziani},
  {Lee}, {Weide}, \& {Xia}}]{Lamb+2010}
{Lamb}, D.~Q., {Couch}, S.~M., {Dubey}, A., {et~al.} 2010, in APS Meeting
  Abstracts, 8010

\bibitem[{{Levermore}(1984)}]{Levermore1984}
{Levermore}, C.~D. 1984, \jqsrt, 31, 149

\bibitem[{{Levermore} \& {Pomraning}(1981)}]{LevermorePomraning1981}
{Levermore}, C.~D., \& {Pomraning}, G.~C. 1981, \apj, 248, 321

\bibitem[{{L\"{o}hner}(1987)}]{Lohner1987}
{L\"{o}hner}, R. 1987, Computer Methods in Applied Mechanics and Engineering,
  61, 323

\bibitem[{{Mac Low} \& {Klessen}(2004)}]{MacLowKlessen2004}
{Mac Low}, M.-M., \& {Klessen}, R.~S. 2004, Reviews of Modern Physics, 76, 125

\bibitem[{{Matzner}(2002)}]{Matzner2002}
{Matzner}, C.~D. 2002, \apj, 566, 302

\bibitem[{{McKee} \& {Ostriker}(2007)}]{McKeeOstriker2007}
{McKee}, C.~F., \& {Ostriker}, E.~C. 2007, \araa, 45, 565

\bibitem[{{Mignone} {et~al.}(2007{\natexlab{a}}){Mignone}, {Bodo}, {Massaglia},
  {Matsakos}, {Tesileanu}, {Zanni}, \& {Ferrari}}]{Mignone2007}
{Mignone}, A., {Bodo}, G., {Massaglia}, S., {et~al.} 2007{\natexlab{a}}, \apjs,
  170, 228

\bibitem[{{Mignone} {et~al.}(2007{\natexlab{b}}){Mignone}, {Bodo}, {Massaglia},
  {Matsakos}, {Tesileanu}, {Zanni}, \& {Ferrari}}]{Mignone+2007}
---. 2007{\natexlab{b}}, \apjs, 170, 228

\bibitem[{{Mihalas} \& {Mihalas}(1984)}]{MihalasMihalas84}
{Mihalas}, D., \& {Mihalas}, B.~W. 1984, {Foundations of radiation
  hydrodynamics} (Oxford University Press)

\bibitem[{{Minerbo}(1978)}]{Minerbo1978}
{Minerbo}, G.~N. 1978, \jqsrt, 20, 541

\bibitem[{{Murray} {et~al.}(2010){Murray}, {Quataert}, \&
  {Thompson}}]{Murray+2010}
{Murray}, N., {Quataert}, E., \& {Thompson}, T.~A. 2010, \apj, 709, 191

\bibitem[{{Murray} {et~al.}(1994){Murray}, {Castor}, {Klein}, \&
  {McKee}}]{Murray+1994}
{Murray}, S.~D., {Castor}, J.~I., {Klein}, R.~I., \& {McKee}, C.~F. 1994, \apj,
  435, 631

\bibitem[{{Nielbock} {et~al.}(2012){Nielbock}, {Launhardt}, {Steinacker},
  {Stutz}, {Balog}, {Beuther}, {Bouwman}, {Henning}, {Hily-Blant},
  {Kainulainen}, {Krause}, {Linz}, {Lippok}, {Ragan}, {Risacher}, \&
  {Schmiedeke}}]{Nielbock+2012}
{Nielbock}, M., {Launhardt}, R., {Steinacker}, J., {et~al.} 2012, \aap, 547,
  A11

\bibitem[{{Offner} {et~al.}(2009){Offner}, {Klein}, {McKee}, \&
  {Krumholz}}]{Offner+2009}
{Offner}, S.~S.~R., {Klein}, R.~I., {McKee}, C.~F., \& {Krumholz}, M.~R. 2009,
  \apj, 703, 131

\bibitem[{{Pascucci} {et~al.}(2004){Pascucci}, {Wolf}, {Steinacker},
  {Dullemond}, {Henning}, {Niccolini}, {Woitke}, \& {Lopez}}]{Pascucci+2004}
{Pascucci}, I., {Wolf}, S., {Steinacker}, J., {et~al.} 2004, \aap, 417, 793

\bibitem[{{Peters} {et~al.}(2011){Peters}, {Banerjee}, {Klessen}, \& {Mac
  Low}}]{Peters+2011}
{Peters}, T., {Banerjee}, R., {Klessen}, R.~S., \& {Mac Low}, M.-M. 2011, \apj,
  729, 72

\bibitem[{{Peters} {et~al.}(2010{\natexlab{a}}){Peters}, {Banerjee}, {Klessen},
  {Mac Low}, {Galv{\'a}n-Madrid}, \& {Keto}}]{Peters+2010a}
{Peters}, T., {Banerjee}, R., {Klessen}, R.~S., {et~al.} 2010{\natexlab{a}},
  \apj, 711, 1017

\bibitem[{{Peters} {et~al.}(2012){Peters}, {Klaassen}, {Mac Low}, {Klessen}, \&
  {Banerjee}}]{Peters+2012}
{Peters}, T., {Klaassen}, P.~D., {Mac Low}, M.-M., {Klessen}, R.~S., \&
  {Banerjee}, R. 2012, \apj, 760, 91

\bibitem[{{Peters} {et~al.}(2014){Peters}, {Klaassen}, {Mac Low}, {Schr{\"o}n},
  {Federrath}, {Smith}, \& {Klessen}}]{Peters+2014}
{Peters}, T., {Klaassen}, P.~D., {Mac Low}, M.-M., {et~al.} 2014, \apj, 788, 14

\bibitem[{{Peters} {et~al.}(2010{\natexlab{b}}){Peters}, {Klessen}, {Mac Low},
  \& {Banerjee}}]{Peters+2010c}
{Peters}, T., {Klessen}, R.~S., {Mac Low}, M.-M., \& {Banerjee}, R.
  2010{\natexlab{b}}, \apj, 725, 134

\bibitem[{{Peters} {et~al.}(2010{\natexlab{c}}){Peters}, {Mac Low}, {Banerjee},
  {Klessen}, \& {Dullemond}}]{Peters+2010b}
{Peters}, T., {Mac Low}, M.-M., {Banerjee}, R., {Klessen}, R.~S., \&
  {Dullemond}, C.~P. 2010{\natexlab{c}}, \apj, 719, 831

\bibitem[{{Price} \& {Bate}(2009)}]{PriceBate2009}
{Price}, D.~J., \& {Bate}, M.~R. 2009, \mnras, 398, 33

\bibitem[{{Rijkhorst} {et~al.}(2006){Rijkhorst}, {Plewa}, {Dubey}, \&
  {Mellema}}]{Rijkhorst+2006}
{Rijkhorst}, E.-J., {Plewa}, T., {Dubey}, A., \& {Mellema}, G. 2006, \aap, 452,
  907

\bibitem[{{Rosdahl} {et~al.}(2013){Rosdahl}, {Blaizot}, {Aubert}, {Stranex}, \&
  {Teyssier}}]{Rosdahl+2013}
{Rosdahl}, J., {Blaizot}, J., {Aubert}, D., {Stranex}, T., \& {Teyssier}, R.
  2013, \mnras, 436, 2188

\bibitem[{Saad \& Schultz(1986)}]{GMRES}
Saad, Y., \& Schultz, M.~H. 1986, SIAM J. Sci. Stat. Comput., 7, 856

\bibitem[{{Safranek-Shrader} {et~al.}(2012){Safranek-Shrader}, {Agarwal},
  {Federrath}, {Dubey}, {Milosavljevi{\'c}}, \& {Bromm}}]{Safranek-Shrader2012}
{Safranek-Shrader}, C., {Agarwal}, M., {Federrath}, C., {et~al.} 2012, \mnras,
  426, 1159

\bibitem[{{Skinner} \& {Ostriker}(2013)}]{SkinnerOstriker2013}
{Skinner}, M.~A., \& {Ostriker}, E.~C. 2013, \apjs, 206, 21

\bibitem[{{Spitzer}(1978)}]{Spitzer1978}
{Spitzer}, L. 1978, {Physical processes in the interstellar medium}

\bibitem[{{Tilley} \& {Pudritz}(2007)}]{TilleyPudritz2007}
{Tilley}, D.~A., \& {Pudritz}, R.~E. 2007, \mnras, 382, 73

\bibitem[{{Turk} {et~al.}(2011){Turk}, {Smith}, {Oishi}, {Skory}, {Skillman},
  {Abel}, \& {Norman}}]{ytpaper}
{Turk}, M.~J., {Smith}, B.~D., {Oishi}, J.~S., {et~al.} 2011, \apjs, 192, 9

\bibitem[{{Turner} \& {Stone}(2001)}]{TurnerStone2001}
{Turner}, N.~J., \& {Stone}, J.~M. 2001, \apjs, 135, 95

\bibitem[{{Vaytet} {et~al.}(2010){Vaytet}, {Audit}, \& {Dubroca}}]{Vaytet+2010}
{Vaytet}, N., {Audit}, E., \& {Dubroca}, B. 2010, in Astronomical Society of
  the Pacific Conference Series, Vol. 429, Numerical Modeling of Space Plasma
  Flows, Astronum-2009, ed. N.~V. {Pogorelov}, E.~{Audit}, \& G.~P. {Zank}, 160

\bibitem[{{Walch} {et~al.}(2012){Walch}, {Whitworth}, {Bisbas}, {W{\"u}nsch},
  \& {Hubber}}]{Walch+2012}
{Walch}, S.~K., {Whitworth}, A.~P., {Bisbas}, T., {W{\"u}nsch}, R., \&
  {Hubber}, D. 2012, \mnras, 427, 625

\bibitem[{{Whitehouse} \& {Bate}(2006)}]{WhitehouseBate2006}
{Whitehouse}, S.~C., \& {Bate}, M.~R. 2006, \mnras, 367, 32

\bibitem[{{Wolfire} \& {Cassinelli}(1986)}]{WolfireCassinelli1986}
{Wolfire}, M.~G., \& {Cassinelli}, J.~P. 1986, \apj, 310, 207

\bibitem[{{Yorke} \& {Sonnhalter}(2002)}]{YorkeSonnhalter2002}
{Yorke}, H.~W., \& {Sonnhalter}, C. 2002, \apj, 569, 846

\end{thebibliography}

\appendix
\section{3T radiation hydrodynamics} \label{sec:appendix}

\FLASH is a code that is supported by a team of developers at the DOE-supported Alliances Center for Astrophysical Thermonuclear Flashes (ASCI) at the University of Chicago\footnote{\url{http://flash.uchicago.edu/}}. Recent development has added significant capabilities for plasma and high-energy-density physics. These included a ``3T'' radiation hydrodynamics solver \citep{Lamb+2010,Fatenejad+2012,Kumar+2011} that evolves an ion fluid, and electron fluid, and a radiation ``fluid''. 3T implies that each fluid has its own temperature, with $T_{\mathrm{ele}} \neq T_{\mathrm{ion}} \neq T_{\mathrm{rad}}$.

These three fluids are coupled in the following way:
\begin{flalign}
\label{eqn:eion_evol}
\partial_t &(\rho e_{\mathrm{ion}}) + \grad \cdot (\rho e_{\mathrm{ion}} \vec{v}) + P_{\mathrm{ion}} \grad \cdot \vec{v} = \\ \nonumber
& \rho \frac{c_{v,\mathrm{ion}}}{\tau_{ei}} \left(T_{\mathrm{ele}} - T_{\mathrm{ion}}\right) \\
\label{eqn:eele_evol}
\partial_t &(\rho e_{\mathrm{ele}}) + \grad \cdot (\rho e_{\mathrm{ele}} \vec{v}) + P_{\mathrm{ele}} \grad \cdot \vec{v} = \\ \nonumber
& \rho \frac{c_{v,\mathrm{ele}}}{\tau_{ei}} \left(T_{\mathrm{ion}} - T_{\mathrm{ele}}\right) - \grad \cdot \vec{q}_{\mathrm{ele}} + Q_{\mathrm{abs}} - Q_{\mathrm{emis}}\\
\partial_t &(\rho e_{\mathrm{rad}}) + \grad \cdot (\rho e_{\mathrm{rad}} \vec{v}) + P_{\mathrm{rad}} \grad \cdot \vec{v} = \\ \nonumber
& \grad \cdot \vec{q}_{\mathrm{rad}} - Q_{\mathrm{abs}} + Q_{\mathrm{emis}}
\end{flalign}

The ion and electron components exchange energy via Coulomb collisions. The electron and radiation components exchange energy through absorption and emission processes. In our case, $Q_{\mathrm{abs}} = \kappa_P \rho c E_r$ and $Q_{\mathrm{emis}} = \kappa_P \rho c a T^4$, where $a$ is the radiation constant. The $\grad \cdot \vec{q}_{\mathrm{ele}}$ and $\grad \cdot \vec{q}_{\mathrm{rad}}$ terms represent the sources or sinks of energy flux for the electron or radiation components.

It has already been shown how the radiation energy sources (stars) couple to the matter fluid. It remains to be shown how the two components of the matter fluid exchange energy in the \FLASH code framework.

The equations to be solved are the specific internal energy updates.
\begin{eqnarray}
\frac{de_{\mathrm{ion}}}{dt} = \frac{c_{v,\mathrm{ion}}}{\tau_{ei}}(T_{\mathrm{ele}} - T_{\mathrm{ion}}) \\
\frac{de_{\mathrm{ele}}}{dt} = \frac{c_{v,\mathrm{ele}}}{\tau_{ei}}(T_{\mathrm{ion}} - T_{\mathrm{ele}})
\end{eqnarray}
The hydrodynamic terms from equations \ref{eqn:eion_evol} and \ref{eqn:eele_evol} are handled separately by the hydro solver (operator splitting). Replacing $de = c_v dT$, we can write the coupled differential equations in terms of temperature:
\begin{eqnarray}
\frac{dT_{\mathrm{ion}}}{dt} = \frac{m}{\tau_{ei}}(T_{\mathrm{ele}} - T_{\mathrm{ion}}) \\
\frac{dT_{\mathrm{ele}}}{dt} = \frac{1}{\tau_{ei}}(T_{\mathrm{ion}} - T_{\mathrm{ele}})
\end{eqnarray}
where $m = c_{v,\mathrm{ele}}/c_{v,\mathrm{ion}}$ is the ratio of the electron and ion specific heats.

This is implemented via a relaxation law, with the temperature ($T = e/c_v$) update
\begin{flalign}
\label{eqn:tion_update}
T_{\mathrm{ion}}^{n+1} =& \left(\frac{T_{\mathrm{ion}}^n + mT_{\mathrm{ele}}^n}{1+m}\right) \\ \nonumber
& - m \left(\frac{T_{\mathrm{ele}}^n + T_{\mathrm{ion}}^n}{1+m}\right) \exp\left[-(1+m)\frac{\Delta t}{\tau_{ei}}\right] \\
\label{eqn:tele_update}
T_{\mathrm{ele}}^{n+1} =& \left(\frac{T_{\mathrm{ion}}^n + mT_{\mathrm{ele}}^n}{1+m}\right) \\ \nonumber
&+ \left(\frac{T_{\mathrm{ele}}^n + T_{\mathrm{ion}}^n}{1+m}\right) \exp\left[-(1+m)\frac{\Delta t}{\tau_{ei}}\right]
\end{flalign}
The equilibration time, $\tau_{ei}$, is given by,
\begin{equation}
\tau_{ei} = \frac{3 k_B^{3/2}}{8 \sqrt{2\pi}e^4} \frac{(m_{\mathrm{ion}}T_{\mathrm{ele}} + m_{\mathrm{ele}}T_{\mathrm{ion}})^{3/2}}{(m_{\mathrm{ele}}m_{\mathrm{ion}})^{1/2}\bar{z}^2 n_{\mathrm{ion}} \ln \Lambda_{ei}},
\end{equation}
where $e$ is the electron charge, $m_{\mathrm{ion}}$ and $m_{\mathrm{ele}}$ are the ion and electron masses, respectively, $\bar{z}$ is the mean ionization level, $n_{\mathrm{ion}}$ is the ion number density, and $\ln \Lambda_{ei}$ is the Coulomb logarithm.

With these updated temperatures, the internal energies are updated:
\begin{eqnarray}
e_{\mathrm{ele}}^{n+1} = e_{\mathrm{ele}}^n + c_{v,\mathrm{ele}}^n(T_{\mathrm{ele}}^{n+1} - T_{\mathrm{ele}}^n) \\
e_{\mathrm{ion}}^{n+1} = e_{\mathrm{ion}}^n + c_{v,\mathrm{ion}}^n(T_{\mathrm{ion}}^{n+1} - T_{\mathrm{ion}}^n)
\end{eqnarray}

%Since our current treatment of radiation hydrodynamics for astrophysical applications does not include ionizing radiation, all material in our simulations is considered neutral. In the limit of zero equilibration time, that is, $\tau_{ei} \rightarrow 0$, the temperature update equations \ref{eqn:tion_update} and \ref{eqn:tele_update} simplify to
%\begin{eqnarray}
%T_{\mathrm{ion}}^{n+1} = \left(\frac{T_{\mathrm{ion}}^n + mT_{\mathrm{ele}}^n}{1+m}\right) \\
%T_{\mathrm{ele}}^{n+1} = \left(\frac{T_{\mathrm{ion}}^n + mT_{\mathrm{ele}}^n}{1+m}\right)
%\end{eqnarray}
%This is equivalent to the fluid being ``instantly'' equilibrated. The energy received by the ``electron'' component is immediately shared with the ``ion'' component, creating a single ``matter'' fluid that behaves just as expected for a neutral medium. The tests described in this paper demonstrate the validity of this approach for neutral gas, while also allowing for the possibility of future experiments of ionized media (e.g. \HII regions) with the electron and ion temperatures not necessarily equal held equal.
%Our approach is also not very computationally expensive, since we do not have to compute the equilibration time $\tau_{ei}$.

\label{lastpage}

\end{document}